\documentclass[12pt,preprint]{aastex}

\newcounter{thefigs}

\newcounter{thetabs}

\newcounter{address}

\def\simless{\mathbin{\lower 3pt\hbox
	{$\,\rlap{\raise 5pt\hbox{$\char'074$}}\mathchar"7218\,$}}} 
\def\simgreat{\mathbin{\lower 3pt\hbox
	{$\,\rlap{\raise 5pt\hbox{$\char'076$}}\mathchar"7218\,$}}} 

\def\ha{H$\alpha$}
\def\hd{H$\delta$}
\def\hb{H$\beta$}
\def\oii{[O{\sc ii}]}
\def\nii{[N{\sc ii}]}

\slugcomment{Submitted to PASJ}

\begin{document}
 
\title{H$\delta$-Selected Galaxies in the Sloan Digital Sky Survey I: The Catalog}

\author{ Tomotsugu Goto\altaffilmark{\ref{CosmicRay},\ref{CarnegieMellon}},
Robert C. Nichol\altaffilmark{\ref{CarnegieMellon}}, Christopher
J. Miller\altaffilmark{\ref{CarnegieMellon}}, Mariangela
Bernardi\altaffilmark{\ref{CarnegieMellon}}, 
Andrew Hopkins\altaffilmark{\ref{Pitt}}, 
Christy Tremonti\altaffilmark{\ref{JHU}}, 
Andrew Connolly\altaffilmark{\ref{Pitt}},
Francisco J. Castander\altaffilmark{\ref{Spain}}, 
J. Brinkmann\altaffilmark{\ref{APO}},
Masataka Fukugita\altaffilmark{\ref{CosmicRay}}, 
Michael Harvanek\altaffilmark{\ref{APO}}, 
\v{Z}eljko Ivezi\'{c}\altaffilmark{\ref{Princeton}},
S.J. Kleinman\altaffilmark{\ref{APO}}, 
Jurek Krzesinski\altaffilmark{\ref{APO},\ref{Suhora}} 
Dan Long\altaffilmark{\ref{APO}}, Jon Loveday \altaffilmark{\ref{Sussex}}, 
Eric H. Neilsen\altaffilmark{\ref{fnal}}, Peter R. Newman\altaffilmark{\ref{APO}},
Atsuko Nitta\altaffilmark{\ref{APO}}, 
Sadanori Okamura\altaffilmark{\ref{Tokyo}}, 
Maki Sekiguchi\altaffilmark{\ref{CosmicRay},}, 
Stephanie A. Snedden\altaffilmark{\ref{APO}}, and 
Mark SubbaRao\altaffilmark{\ref{Chicago}} }

\setcounter{address}{1}
\altaffiltext{\theaddress}{ 
\stepcounter{address} Institute for Cosmic Ray
Research, Univ. of Tokyo, Kashiwanoha, Kashiwa, Chiba 2770882, Japan
\label{CosmicRay}}
\altaffiltext{\theaddress}{
\stepcounter{address}
Department of Physics, Carnegie Mellon University, 
5000 Forbes Avenue, Pittsburgh, PA 15213-3890 
\label{CarnegieMellon}}
\altaffiltext{\theaddress}{
\stepcounter{address}
Department of Physics and Astronomy, University of Pittsburgh, 3941 O'Hara St., Pittsburgh, PA 15260
\label{Pitt}}
\altaffiltext{\theaddress}{
\stepcounter{address}
Department of Physics and Astronomy, The Johns Hopkins University, 3400 North Charles Street, Baltimore, MD 21218-2686, USA
\label{JHU}}
\altaffiltext{\theaddress}{
\stepcounter{address}
Institut d'Estudis Espacials de Catalunya/CSIC, Gran Capit\`a 2-4, 
08034 Barcelona, Spain
\label{Spain}}
\altaffiltext{\theaddress}{
\stepcounter{address}
Princeton University Observatory, Princeton,
NJ 08544
\label{Princeton}}
\altaffiltext{\theaddress}{
\stepcounter{address}
Apache Point Observatory,
     2001 Apache Point Road,
     P.O. Box 59, Sunspot, NM 88349-0059, USA
\label{APO}}
\altaffiltext{\theaddress}{
\stepcounter{address}
Mt. Suhora Observatory, Cracow Pedagogical University, ul. Podchorazych 2, 30-084 Cracow, Poland
\label{Suhora}}
\altaffiltext{\theaddress}{
\stepcounter{address}
Sussex Astronomy Centre,
University of Sussex,
Falmer, Brighton BN1 9QJ, UK
\label{Sussex}}
\altaffiltext{\theaddress}{
\stepcounter{address}
Fermi National Accelerator Laboratory, P.O. Box 500,
Batavia, IL 60510
\label{fnal}}
\altaffiltext{\theaddress}{
\stepcounter{address}
Department of Astronomy and Research Center for the
   Early Universe, School of Science, University of Tokyo, Tokyo
   113-0033, Japan
\label{Tokyo}}
\altaffiltext{\theaddress}{
\stepcounter{address}
Department of Astronomy and Astrophysics, University of Chicago, 5640 South Ellis Avenue, Chicago, IL 6063
\label{Chicago}}

\clearpage

\newpage

\begin{abstract}
We present here a new and homogeneous sample of 3340 galaxies selected from
the Sloan Digital Sky Survey (SDSS) based solely on the observed strength of
their \hd\, Hydrogen Balmer absorption line. The presence of a strong \hd\,
line within the spectrum of a galaxy indicates that the galaxy has under--gone
a significant change in its star formation history within the last
Gigayear. Therefore, such galaxies have received considerable attention in
recent years, as they provide an opportunity to study galaxy evolution {\it in
action}. These galaxies are commonly known as ``post--starburst'', ``E+A'',
``k+a'' and \hd--strong galaxies, and the study of these galaxies has been
severely hampered by the lack of a large, statistical sample of such
galaxies. In this paper, we rectify this problem by selecting a sample of
galaxies which possess an absorption \hd\, equivalent width of ${\rm
EW(H\delta_{\rm max})} - \Delta {\rm EW(H\delta_{\rm max})} >4$\AA from 106682
galaxies in the SDSS.  We have performed extensive tests on our catalog
including comparing different methodologies of measuring the \hd\, absorption
lines and studying the effects of stellar absorption, dust extinction and
emission--filling on our measurements. We have determined the external error
on our \hd\, measurements using duplicate observations of 11538 galaxies in
the SDSS. The measured abundance of our \hd--selected (HDS) galaxies is
$2.6\pm0.1$\% of all galaxies within a volume--limited sample of $0.05<z<0.1$
and M($r^*)<-20.5$, which is consistent with previous studies of such galaxies
in the literature. We find that only 25 of our HDS galaxies in this
volume--limited sample ($3.5\pm0.7$\%) show no, or little, evidence for \oii\,
and \ha\, emission lines, thus indicating that true E+A (or k+a) galaxies (as
originally defined by Dressler \& Gunn) are extremely rare objects at low
redshift, {\it i.e.}, only $0.09\pm0.02$\% of all galaxies in this
volume--limited sample are true E+A galaxies. In contrast, $89\pm5$\% of our
HDS galaxies in the volume--limited sample have significant detections of the
\oii\, and \ha\, emission lines. Of these, only 131 galaxies are robustly
classified as Active Galactic Nuclei (AGNs) and therefore, a majority of
these emission line HDS galaxies are star--forming galaxies.  We find that
$52\pm12$\% (27/52) of galaxies in our volume--limited HDS sample that possess
no detectable \oii\, emission, do however possess detectable \ha\, emission
lines. These galaxies may be dusty star--forming galaxies.  We provide the
community with this new catalog of \hd--selected galaxies to aid in the
understanding of these galaxies, via detailed follow--up observations, as well
as providing a low redshift sample for comparison with higher redshift studies
of HDS galaxies.  We will study the global properties of these galaxies in
future papers.
\end{abstract}

\section{Introduction}\label{intro}

The presence of a strong \hd\, absorption line (equivalent width of $\simgreat
5$\AA) in the spectrum of a galaxy is an indication that the spectral energy
distribution of that galaxy is dominated by A stars. Models of galaxy
evolution indicate that such a strong \hd\, line (in the spectrum of a galaxy)
can only be reproduced using models that include a recent burst of star
formation, followed by passive evolution, as any on--going star--formation in
the galaxy would hide the \hd\, absorption line due to emission--filling (of
the \hd\, line) and the dominance of hot O and B stars, which have
intrinsically weaker \hd\, absorption than A stars (see, for example, Balogh
et al. 1999; Poggianti et al. 1999). Therefore the existence of a strong \hd\,
absorption line in the spectrum of a galaxy suggests that the galaxy has
under--gone a recent transformation in its star--formation history. In the
literature, such galaxies are called ``post--starburst'', ``E+A'', k+a, and
H$\delta$--strong galaxies and the exact physical mechanism(s) responsible for
the abrupt change in the star formation history of such galaxies remains
unclear. These galaxies have received much attention as they provide an
opportunity to study galaxy evolution {\it ``in action''}.

\hd--strong galaxies were first discovered by Dressler \& Gunn (1983,
1992) in their spectroscopic study of galaxies in distant, rich clusters of
galaxies.  They discovered cluster galaxies that contained strong Balmer
absorption lines but with no detectable \oii\, emission lines. They named such
galaxies ``E+A'', as their spectra resembled the superposition of an
elliptical galaxy spectrum and A star spectrum. Therefore, E+A (or k+a) galaxies were
originally thought to be a cluster--specific phenomenon and several physical
mechanisms have been proposed to explain such galaxies.  For example,
ram--pressure stripping of the intra--stellar gas by a hot, intra--cluster
medium, which eventually leads to the termination of star formation once all
the gas in the galaxy has been removed, or used up (Gunn \& Gott 1972; Farouki
\& Shapiro 1980; Kent 1981; Abadi, Moore \& Bower 1999; Fujita \& Nagashima
1999; Quilis, Moore \& Bower 2000). Alternative mechanisms include high--speed
galaxy--galaxy interactions in clusters (Moore et al. 1996, 1999) and
interactions with the gravitational potential well of the cluster (Byrd \&
Valtonen 1990; Valluri 1993; Bekki, Shioya \& Couch 2001).

To test such hypotheses, Zabludoff et al. (1996) performed a search for
E+A (or k+a) galaxies in the Las Campanas Redshift Survey
(LCRS; Shectman et al. 1996) and found that only 21 of the 11113 LCRS galaxies
they studied, satisfied their criteria for a E+A galaxy. This work
clearly demonstrates the rarity of such galaxies at low redshift. Furthermore,
Zabludoff et al. (1996) found that 75\% of their selected galaxies
reside in the field, rather than the cores of rich clusters. This conclusion
was confirmed by Balogh et al. (1999), who also performed a search for
\hd--strong galaxies in the redshift surveys of the Canadian Network for
Observational Cosmology (CNOC; Yee, Ellingson, \& Carlberg 1996), and found
that the fraction of such galaxies in clusters was consistent with
that in the field. Alternatively, the study of Dressler et al. (1999) found an
order--of--magnitude increase in the abundance of E+A (or k+a) galaxies in
distant clusters compared to the field (see also Castander et al. 2001).
Taken together, these studies suggest that the physical interpretation of
\hd--strong galaxies is more complicated than originally envisaged, with
the possibility that different physical mechanisms are important in different
environment, {\it e.g.}, 5 of the 21 E+A galaxies discovered by
Zabludoff et al. (1996) show signs of tidal features, indicative of
galaxy--galaxy interactions or mergers. Furthermore, redshift evolution might
be an important factor in the differences seen between these surveys.

In addition to studying the environment of \hd--strong galaxies, several
authors have focused on understanding the morphology and dust content of these
galaxies. This has been driven by the fact that on--going star formation in
post--starburst galaxies could be hidden by dust obscuration (See Poggianti \&
Barbaro 1997; Poggianti et al. 1999; Bekki et al. 2001 for more
discussion). In fact, Smail et al (1999) discovered examples of such galaxies
using infrared (IR) and radio observations of galaxies in distant
clusters. They discovered five post--starburst galaxies (based on their
optical spectra) that showed evidence for dust--lanes in their IR morphology,
as well as radio emission consistent with on--going star formation. However,
radio observations of the Zabludoff et al. (1996) sample of nearby
E+A galaxies indicates that a majority of these galaxies are not
dust--enshrouded starburst galaxies. For example, Miller \& Owen (2001) only
detected radio emission from 2 of the 15 E+A galaxies they
observed, and the derived star--formation rates (SFRs) were consistent with
quiescent star formation and thus much lower than those observed for the
dust--enshrouded starburst galaxies of Smail et al (1999). Chang et al. (2001)
also did not detect radio emission from 5 E+A galaxies they
observed from the Zabludoff et al. (1996) sample and concluded that these
galaxies were not dust--enshrouded starbursts. In summary, these studies
demonstrate that some E+A galaxies have dust--enshrouded star
formation, but the fraction remains ill--determined. Furthermore, it is
unclear how these different sub--classes of galaxies are related, and if there
are any environmental and evolutionary processes at play.

The interpretation of \hd--strong galaxies (E+A, k+a, {\it etc.}) suffers from
small number statistics and systematic differences in the selection and
definition of such galaxies between the different surveys constructed to date.
Therefore, many of the difficulties associated with understanding the physical
nature of these galaxies could be solved through the study of a large,
homogeneous sample of \hd\, galaxies. In this paper, we present such a sample
derived from the Sloan Digital Sky Survey (SDSS; York et al. 2000). The
advantage of this sample, over previous work, is the quality and quantity of
both the photometric and spectroscopic data, as well as the homogeneous
selection of SDSS galaxies which covers a wide range of local environments.

We present in this paper a sample of galaxies that have been selected based
solely on the observed strength of their \hd\, absorption line. Our selection
is thus inclusive, containing many of the sub--classes of galaxies discussed
in the literature until now, {\it e.g.}, ``E+A'' or ``k+a'' galaxies
(Zabludoff et al. 1996; Dressler et al. 1999), post--starburst galaxies,
dust--enshrouded starburst galaxies (Smail et al. 1999), \hd--strong galaxies
(Couch \& Sharples 1987) and the different subsamples of galaxies ({\it i.e.},
e(a), A+em) discussed by Poggianti et al. (1999) and Balogh et
al. (1999). Therefore, to avoid confusion with other samples in the
literature, we simply call this sample of SDSS galaxies; ``\hd--selected''
(HDS) galaxies.

In this paper, we present the details of our selection and leave the
investigation and interpretation of these HDS galaxies to subsequent papers.
We publish our sample of HDS galaxies to help the community construct larger
samples of such galaxies, which are critically needed to advance our
understanding of these galaxies, as well as aiding in the planning of
follow--up observations and comparisons with higher redshift studies of such
galaxies.

In Section \ref{data}, we present a brief discussion of the SDSS and the data
used in this paper. In Section \ref{line}, we discuss our techniques for
measuring the \hd\, absorption line and present comparisons between the
different methodologies used to measure this line. In Section \ref{catalog},
we discuss the criteria used to select of our HDS sample of galaxies and
present data on 3340 such galaxies in our catalog.  In Section
\ref{discussion}, we compare our sample of galaxies with those in the
literature. A more detailed analysis of the properties of our HDS galaxies
will be discussed in subsequent papers. The cosmological parameters used
throughout this paper are ${\rm H_0}$=75 km s$^{-1}$ Mpc$^{-1}$,
$\Omega_m=0.3$ and $\Omega_{\Lambda}=0.7$.

\section{The SDSS Data}
\label{data}

In this Section, we briefly describe the spectroscopic part of the SDSS. As
discussed in York et al. (2000), the SDSS plans to obtain spectra for
$\simeq10^6$ galaxies to a magnitude limit of $r^*=17.7$ (the ``Main'' galaxy
sample; Strauss et al. 2002), $\simeq10^5$ Luminous Red Galaxies (LRG;
Eisenstein et al. 2001) and $\simeq10^5$ quasars (Richards et al. 2002).  The
reader is referred to Fukugita et al. (1996), Gunn et al. (1998), Lupton et
al. (1999, 2001), York et al. (2000), Hogg et al. (2001), Pier et al. (2002),
Stoughton et al. (2002), Smith et al. (2002) and Blanton et al. (2002a) for
more details of the SDSS data and survey.

The SDSS spectra are obtained using two fiber-fed spectrographs (each with 320
fibers), with each fiber sub-tending 3 arcseconds on the sky. The wavelength
coverage of the spectrographs is 3800\AA\ to 9200\AA, with a spectral
resolution of 1800. The data from these spectrographs is automatically reduced
to produce flux and wavelength--calibrated spectra (SPECTRO2D data analysis
pipeline).

The SDSS spectra are then analyzed via the SDSS SPECTRO1D data processing
pipeline to obtain a host of measured quantities for each spectrum (see
Stoughton et al. 2002; Frieman et al., in prep, for further details). For
example, SPECTRO1D determines the redshift of the spectrum both from
absorption lines (via cross-correlation; Heavens 1993), and emission lines
(via a wavelet--based peak--finding algorithm; see Frieman et al., in prep).
Once the redshift is known, SPECTRO1D estimates the continuum emission at each
pixel using the median value seen in a sliding box of 100 pixels centered on
that pixel.  Emission and absorption lines are then measured automatically by
fitting of a Gaussian, above the best--fit continuum, at the redshifted
rest--wavelength of expected lines. Multiple Gaussians are fit simultaneously
for potential blends of lines ({\it i.e.}, \ha\, and \nii\, lines). SPECTRO1D
therefore provides an estimate of the equivalent width (EW), continuum, rest
wavelength, identification, goodness--of--fit ($\chi^2$), height and sigma
(and the associated statistical errors on these quantities) for all the major
emission/absorption lines in these spectra. These measurements are done
regardless of whether the line has been detected or not.  For this work, we
have used data from rerun 15 of the SPECTRO1D analysis pipeline, which is
based on version 4.9 of the SPECTRO2D analysis pipeline (see Frieman et al. in
prep for details of these pipelines).

For the sample of HDS galaxies presented in this paper, we begin with a sample
of the SDSS galaxies that satisfy the following selection criteria:
\begin{enumerate}
\item{Spectroscopically--confirmed by SPECTRO1D to be a galaxy;}
\item{Possess a redshift confidence of $\geq0.7$;}
\item{An average spectroscopic signal--to--noise of $>5$ per pixel in
     the SDSS photometric $g$ passband;}
 \item {z$\geq$0.05, to minimize aperture effects as discussed in Zaritsky, Zabuldoff \& Willick (1995) and Gomez et al. (2003).}
\end{enumerate}

\begin{figure}
\plotone{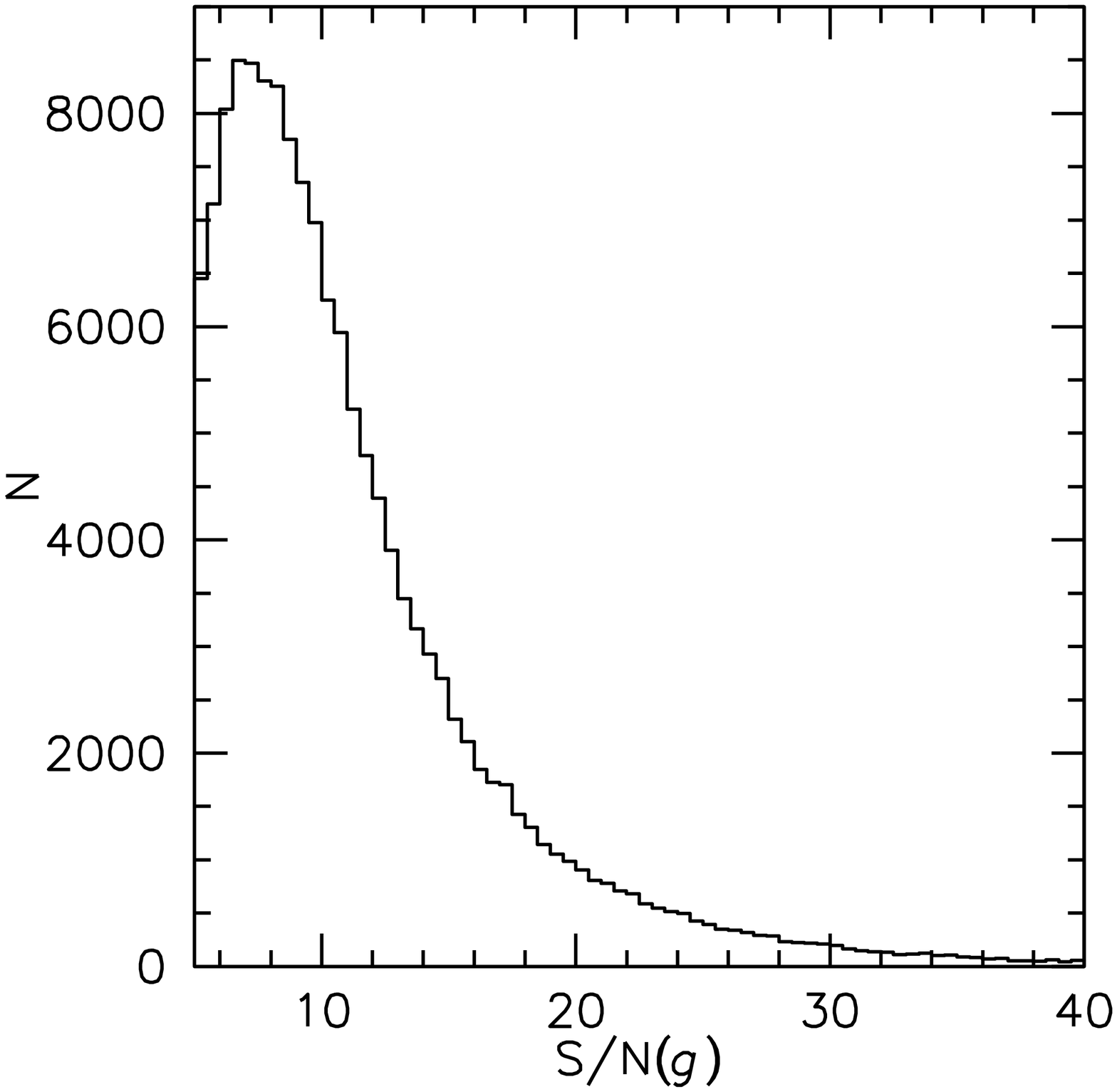}
\caption{
\label{fig:sng}
The distribution of signal--to--noise for spectra used in this
analysis (see Section \ref{data}). The signal--to--noise presented here is the
average signal--to--noise per pixel over the wavelength range defined by the
SDSS photometric $g$ passband. The median signal--to--noise value is
8.3. Galaxies with signal--to--noise less than 5 were not used in our study.
 }
\end{figure}

The reader is referred to Stoughton et al. (2002) for further details on all
these SDSS quantities and how they are determined.  After removing duplicate
observations of the same galaxy (11538 galaxies in total; see Section
\ref{error}), 106682 galaxies satisfy these criteria, up to and including
spectroscopic plate 804 (observed on a Modified Julian Date of 52266 or
12/23/01; see Stoughton et al. 2002). Of these 106682 galaxies, it was only
possible to measure the \hd\, line for 95479 galaxies (see Section
\ref{Hdelta} below) due to masked pixels at or near the \hd\, line. In Figure
\ref{fig:sng}, we present the distribution of signal--to--noise ratios for all
106682 spectra (the median value of this distribution is 8.3).

Throughout this analysis, we have used the ``smeared'' SDSS spectra, which
improves the overall spectrophotometric calibration of these data by
accounting for light missed from the 3'' fibers due to seeing and
atmospheric refraction (see Gomez et al. 2003; Stoughton et al. 2002 for
observational detail). Unfortunately, this smearing correction can
systematically bias the observed flux of any emission and absorption lines in
the spectrum, as the correction is only applied to the continuum. As shown by
Hopkins et al. in prep., this is only a $\simeq10\%$ effect on the flux of
spectral lines, compared to using spectral data without the smearing
correction applied. Furthermore, the equivalent width of our lines is almost
unaffected by this smearing correction as, by definition, they are computed
relative to the height of the continuum.

\section{Spectral Line Measurements}\label{line}
\subsection{H$\delta$\ Equivalent Width}\label{Hdelta}

In this Section, we discuss the measurement of the equivalent width (EW) of
the \hd\, absorption line in the SDSS galaxy spectra described in Section
\ref{data}. The presence of a strong \hd\, absorption line in a galaxy
spectrum indicates that the stellar population of the galaxy contains a
significant fraction of A stars, which must have formed within the last
Gigayear (see Section \ref{intro}).  The \hd\, line is preferred over other
Hydrogen Balmer lines ({\it e.g.}, H$\epsilon$, H$\zeta$, H$\gamma$, H$\beta$)
because the line is isolated from other emission and absorption lines, as well
as strong continuum features in the galaxy spectrum ({\it e.g.},
D4000). Furthermore, the higher order Balmer lines (H$\gamma$ and H$\beta$)
can suffer from significant emission--filling (see Section \ref{emfilling}),
while the lower order lines (H$\epsilon$ and H$\zeta$) are low
signal--to--noise in the SDSS spectra.

In previous studies, several different methods have been employed to measure
the \hd\, line, or select post--starburst galaxies. For example, Zabludoff et
al. (1996) used the average EW of the \hb\ , H$\gamma$ and H$\delta$ lines to
select E+A galaxies. Alternatively, Dressler et al. (1999) and Poggianti et
al. (1999) interactively fit Gaussian profiles to the H$\delta$ line. Finally,
Abraham et al. (1996), Balogh et al. (1999) and Miller \& Owen (2002)
performed a non--parametric analysis of their galaxy spectra, which involved
summing the flux in a narrow wavelength window centered on each of the \hd\,
line to determine the EW of the line. Castander et al. (2001) used an
innovative PCA and wavelet analysis of spectra to select E+A galaxies.  Each
of these methods have different advantages and dis--advantages. For example,
fitting a Gaussian to the \hd\, line is optimal for high signal--to--noise
spectra, but can be prone to erroneous results when fit blindly to low
signal--to--noise data or a weak absorption lines (such problems can be
avoided if Gaussians are fit interactively; see Dressler et al. 1999;
Poggianti et al. 1999).  In light of the potential systematic differences
between the different methods of measuring the \hd\, line, we have
investigated the relative merits of the two main approaches in the literature
-- fitting a Gaussian and summing the flux in a narrow wavelength window --
for determining the EW of the \hd\, line for the signal--to--noise,
resolution, and size of the SDSS spectral dataset used in this paper.

First, we investigate the optimal method for computing the EW of the \hd\,
line from the SDSS spectra using the non--parametric methodology outlined in
Abraham et al. (1996) and Balogh et al. (1999), {\it i.e.}, summing the flux
within narrow wavelength windows centered on and off the \hd\ absorption
line. We estimate the continuum flux via linear interpolation between two
wavelength windows placed either side of the \hd\, line (4030\AA\ to 4082\AA\
and 4122\AA\ to 4170\AA). We used the same wavelength windows as in Abraham et
al. (1996) and Balogh et al. (1999) for estimating the continuum because they
are devoid of any strong emission and absorption features, and the continuum
is relatively smooth within these wavelength ranges. Also, these windows are
close to the \hd\, line without being contaminated by the \hd\, line emission.
When fitting the continuum flux level, the flux in each pixel was weighted by
the inverse square of the error on the flux in that pixel.  After the initial
fit to the continuum, we re--iterate the fit once by rejecting $3\sigma$
outliers to the original continuum fit.  This guards against noise spikes in
the surrounding continuum.

The rest--frame EW of the \hd\, line was calculated by summing the ratio of
the flux in each pixel of the spectrum, over the estimated continuum flux in
that pixel based on our linear interpolation.  For this summation, we
investigated two different wavelength windows for the \hd\, line; 4088\AA\ to
4116\AA, which is the same as the wavelength range used by Balogh et
al. (1999)$\footnote{We note that Table 1 of Balogh et al. (1999) has a
typographical error. The authors used the wavelength range of 4088\AA\ to
4116\AA\ to measure their \hd\, EWs instead of 4082\AA\ to 4122\AA\ as quoted
in the paper.}$ and 4082\AA\ to 4122\AA, which is the wider range of Abraham
et al. (1996). We summarize the wavelength ranges used to measure the \hd\,
EWs in Table \ref{tab:wavelength}.

\begin{figure}
\plotone{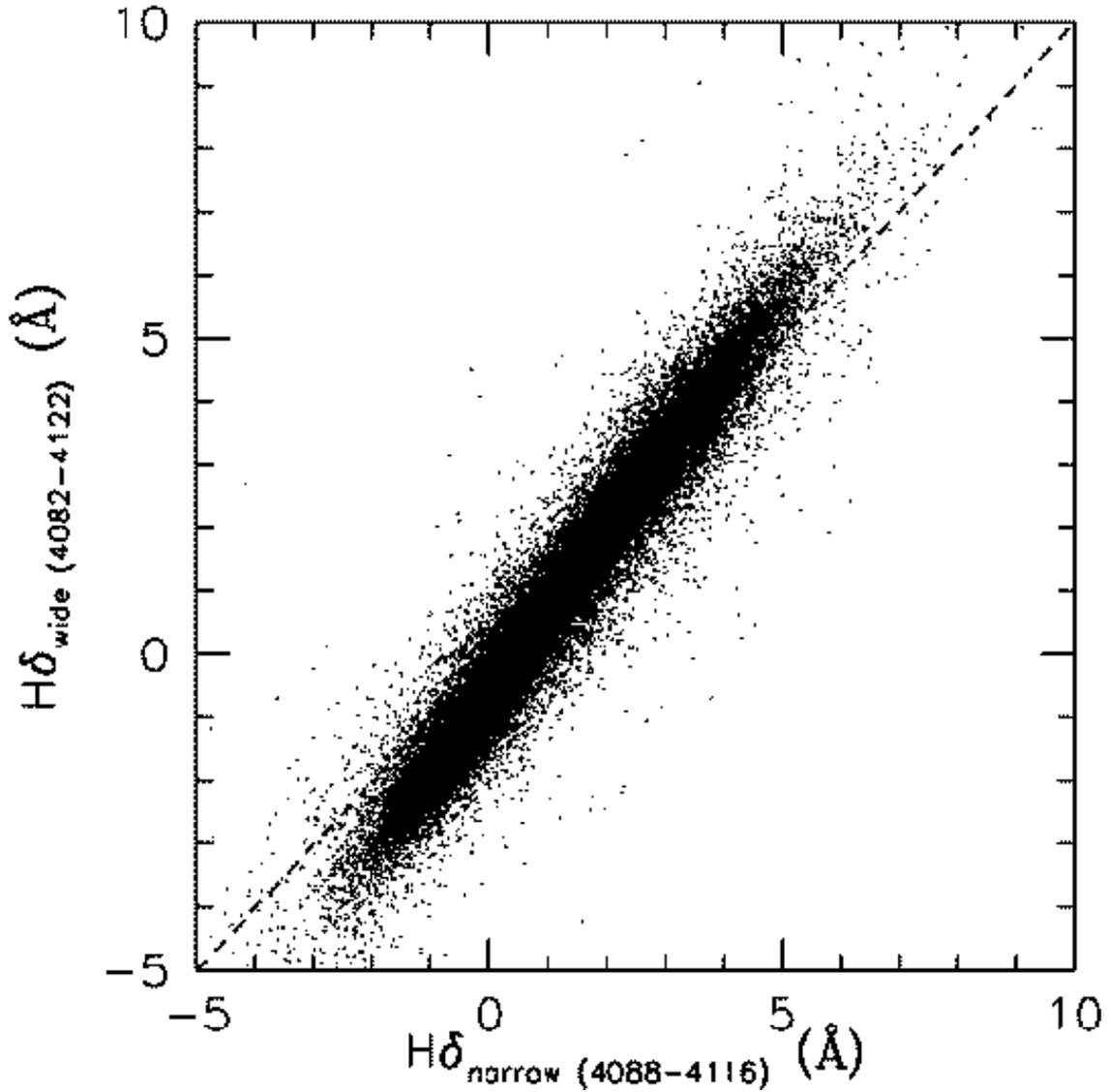}
\caption{
\label{fig:hd_narrow_wide}
The \hd\, EW (\AA) as measured in two different wavelength windows, {\it i.e.},
the {\it wide} window of Abrahams et al. (1996) and the {\it narrow} window
of Balogh et al. (1999). The expected one--to--one line is plotted to help
guide the eye. For the work presented in this paper, absorption lines have a
positive EW values and emission lines have negative EW values. }
\end{figure}

In Figure \ref{fig:hd_narrow_wide}, we compare the two non--parametric
measurements of \hd\ ({\it i.e.}, using the narrow and wide wavelength
windows), and find, as expected, a strong linear relationship between the two
measurements: The scatter about the best fit linear relationship to these
measurements is Gaussian with $\sigma=0.29$\AA. However, there are systematic
differences between the two measurements which are correlated to the intrinsic
width of the \hd\, line. For example, for large EWs of \hd, we find that the
wider wavelength window has a larger value than the narrower window. This is
because the 28\AA\ window is too small to capture the wings of a strong \hd\,
line and thus a wider window is needed. Alternatively, for smaller EWs, the
narrower 28\AA\ window is better as the larger window of 40\AA\ is more
affected by noise as it contains more of the continuum flux than the narrow
window. This is seen in Figure \ref{fig:hd_narrow_wide} where the larger
window systematically over--estimates the EW of the weaker \hd\, lines. 

As a compromise, we have empirically determined that the best methodology for
our analysis is to always select the larger of the two \hd\, EW measurements
(this was discovered by visually inspecting many of the spectra and their
various \hd\, measurements). This is a crude adaptive approach of selecting
the size of the window based on the intrinsic strength of the \hd\, line. In
fact, we find that 20.2\% of our \hd--selected galaxies (see Section
\ref{catalog}) were selected based on the \hd\, measurement in the larger
wavelength window.
  Therefore, for the analysis presented in this
paper, we use H$\delta_{\rm max}$, which is the maximum of the two
non--parametric measurements discussed above.

\begin{figure}
\plotone{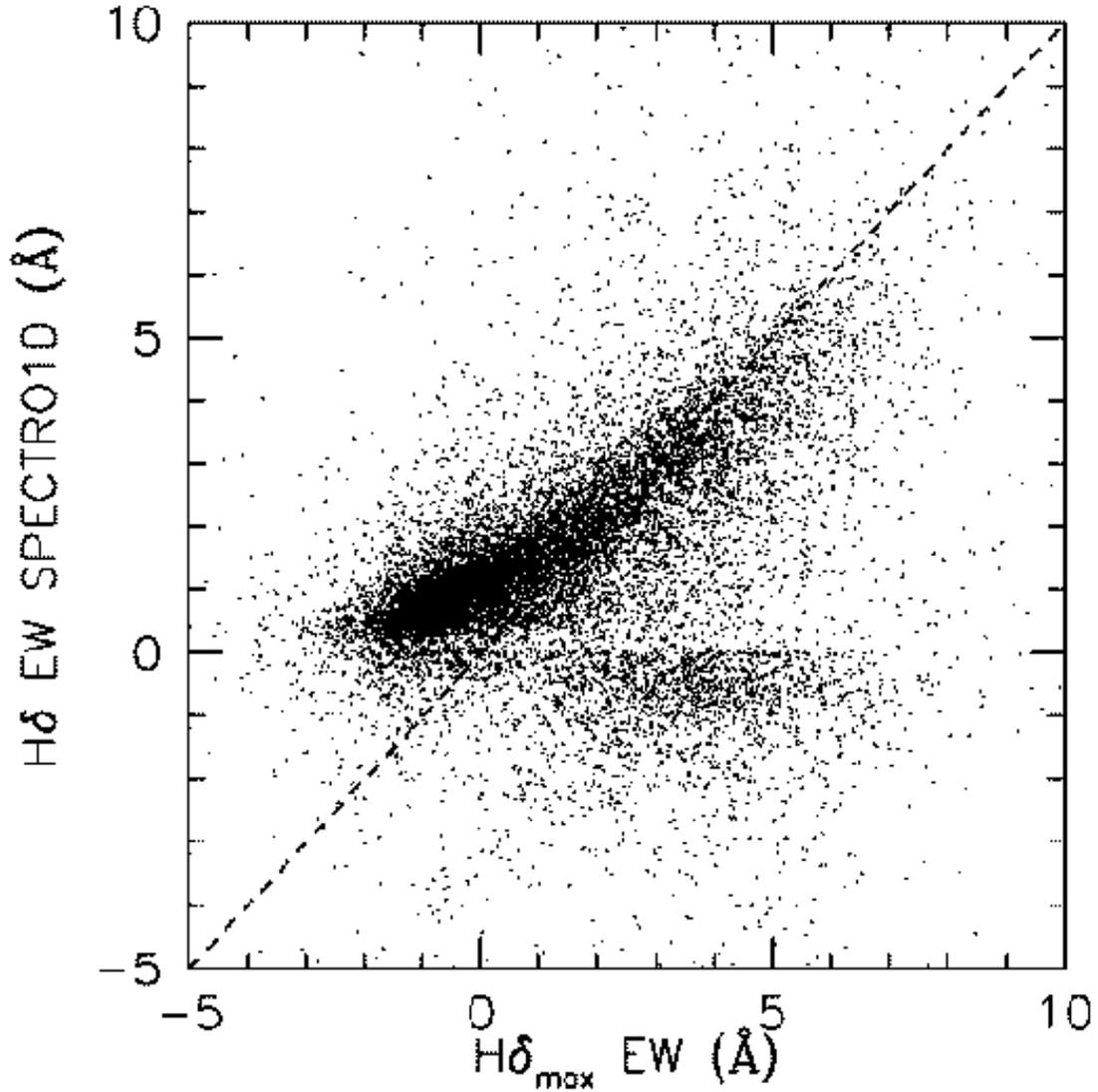}
\caption{
\label{fig:hd_max_1d} 
A comparison of the \hd\, EW as measured automatically by the SDSS SPECTRO1D
spectroscopic pipeline (a Gaussian fit to the \hd\, line) and the
non--parametric summation technique discussed in this paper and presented in
Figure \ref{fig:hd_narrow_wide}.  The one--to--one line is shown to guide the
eye.  In our work, absorption lines have positive EW values and emission lines
have negative EW values.  }
\end{figure}

\begin{figure}
\plotone{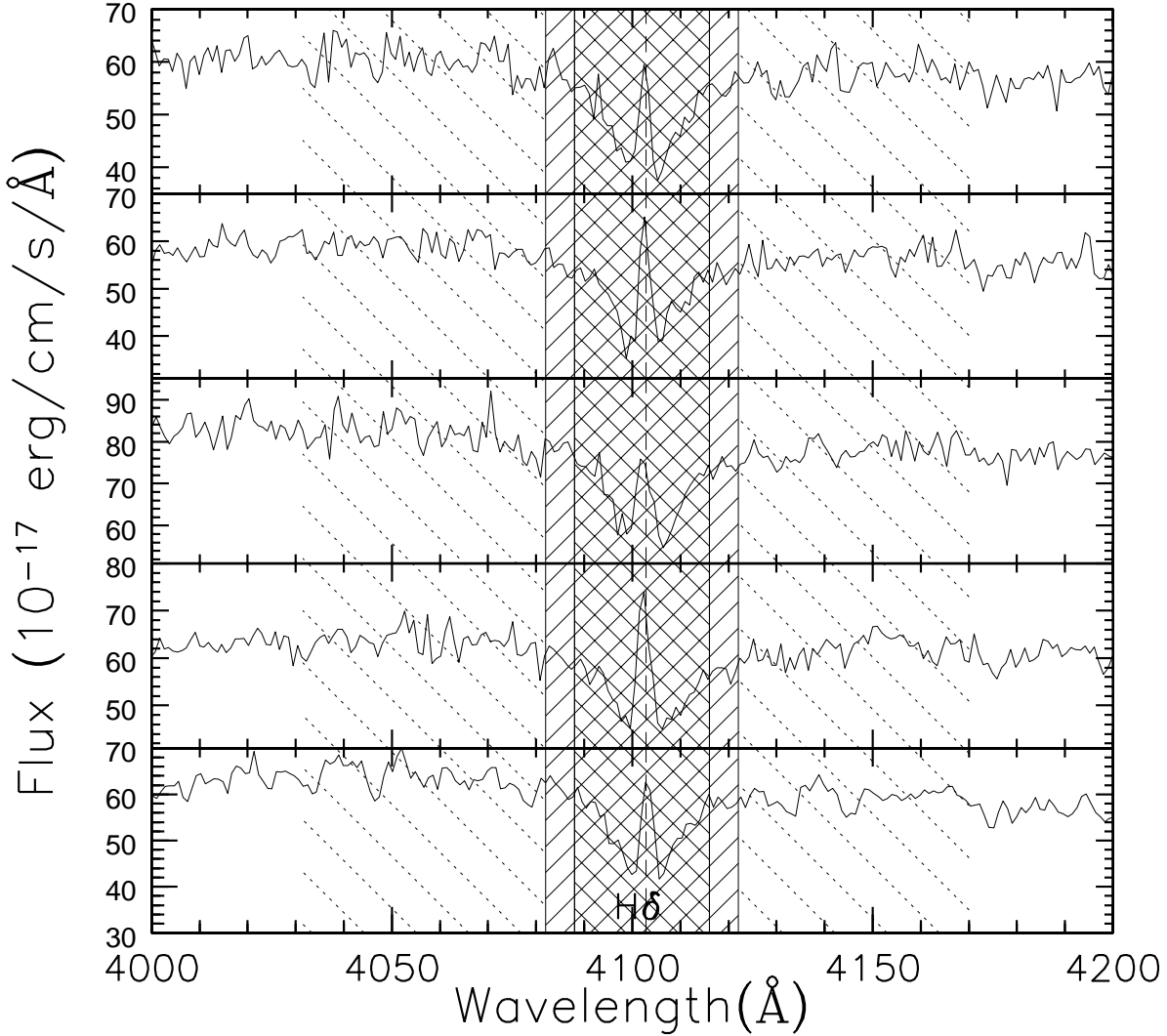}
\caption{
\label{fig:spec-0293-51689-144.ps}
Five typical examples of SDSS spectra with H$\delta$ emission filling.
In such cases it is difficult to fit the H$\delta$ absorption emission
with a single Gaussian due to a centrally peaked emission. The double
hashed region of this figure, centered on the \hd\, line, represents
the narrow wavelength window used to measure the EW of \hd\, as
explained in Section \ref{Hdelta}. The slightly wider hashed region,
again centered on the \hd\, line, represented the wide wavelength
window used to measure the \hd\, line (see Section
\ref{Hdelta}). Finally, the dashed regions, one on each side of the hashed
regions, represents the wavelength regions used to estimate the
continuum flux. See also Table 1 for details of the wavelength windows
used on measuring the \hd\, line.}
\end{figure}

In Figure \ref{fig:hd_max_1d}, we now compare the H$\delta_{\rm max}$
measurement discussed above to the automatic Gaussian fits to the \hd\, line
from the SDSS SPECTRO1D analysis of the spectra. As expected, the two methods
give similar results for the EW of the \hd\, line for the largest
EWs. However, there are significant differences, as seen in Figure
\ref{fig:hd_max_1d}, between these two methodologies.  First, there are many
galaxies with a negative EW (emission) as measured by SPECTRO1D, but possess a
(large) positive EW (absorption) using the non--parametric method. These cases
are caused by emission--filling, {\it i.e.}, a small amount of \hd\, emission
at the bottom of the \hd\ absorption line (see Section \ref{emfilling}). This
results in SPECTRO1D fitting the Gaussian to the central emission line, thus
producing a negative EW. On the other hand, the non--parametric method simply
sums all the flux in the region averaging over the emission and still
producing a positive EW. In Figure \ref{fig:spec-0293-51689-144.ps}, we
present five typical examples of this phenomenon.
 
\begin{figure}
\plotone{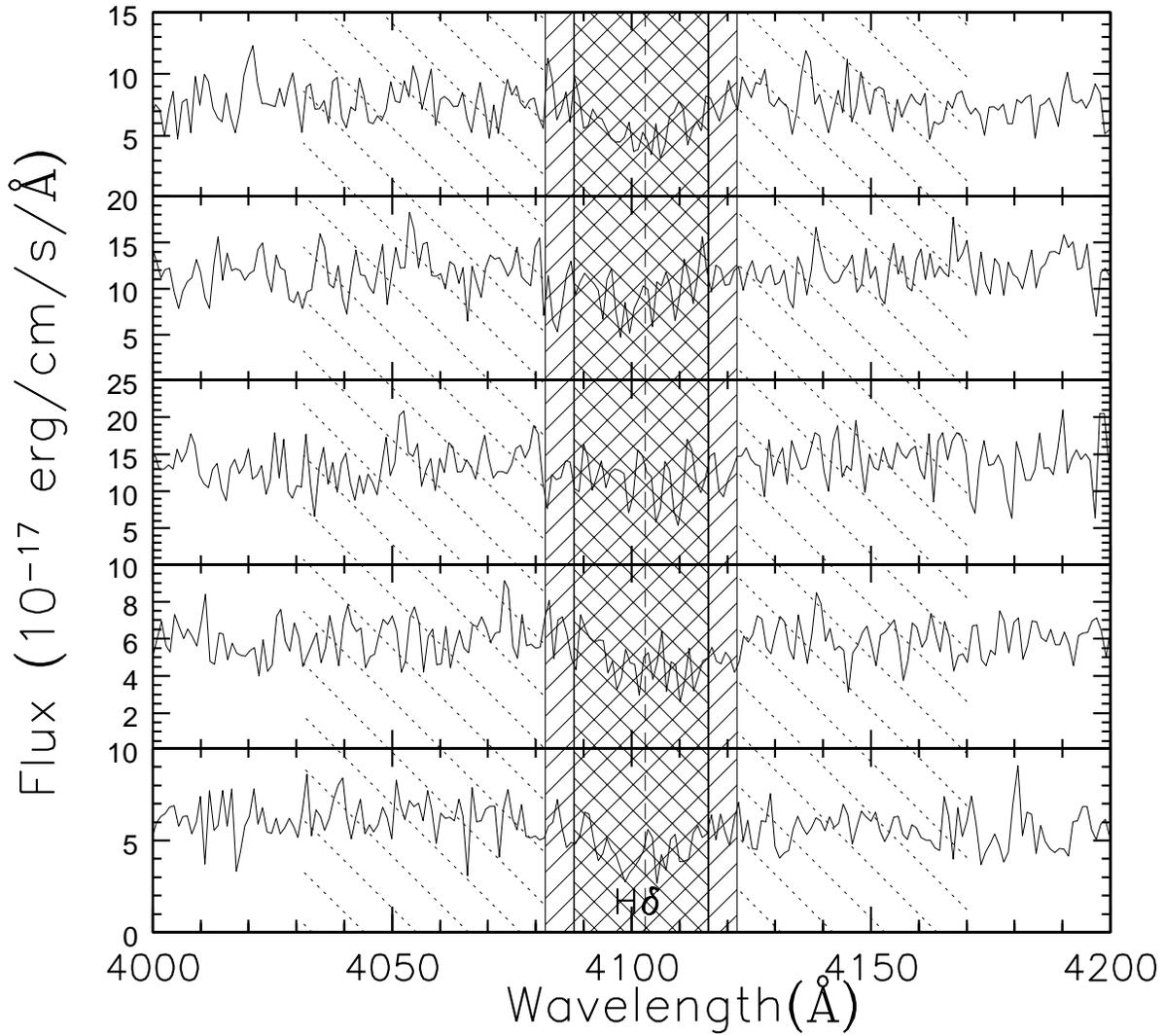}
\caption{
\label{fig:spec-0415-51810-304.ps}
Five examples of noisy spectra where the SDSS SPECTRO1D pipeline has fit
 a broad absorption line, thus overestimating the \hd\, EW. 
The hashed regions are the same as presented and discussed in Figure 
\ref{fig:spec-0293-51689-144.ps}.
}
\end{figure}

Another noticeable difference in Figure \ref{fig:hd_max_1d} is the deviation
from the one--to--one relation for \hd\, EWs near zero, {\it i.e.}, as the
\hd\, line becomes weak, it is buried in the noise of the continuum making it
difficult to automatically fit a Gaussian to the line. In such cases,
SPECTRO1D tends to over--estimate the EW of the \hd\, line because it
preferentially fits a broad, shallow Gaussian to the noise in the
spectrum. Typical examples of this problem are shown in Figure
\ref{fig:spec-0415-51810-304.ps}. We conclude from our study of the SDSS
spectra that the non--parametric techniques of Abraham et al. (1996) and
Balogh et al. (1999) are preferred to the automatic Gaussian fits of
SPECTRO1D, especially for the lower signal--to--noise SDSS spectra which are a
majority in our sample (see Figure \ref{fig:sng}). We note that many of the
problems associated with the automatic Gaussian fitting of SPECTRO1D can be
avoided by fitting Gaussians interactively. However, this is not practical for
large datasets such as the SDSS.

\subsection{\oii\, and \ha\, Equivalent Widths}\label{oiiha}

In addition to estimating the EW of the \hd\, line, we have used our
flux--summing technique to estimate the rest frame equivalent widths of both
the \oii\, and \ha\, emission lines. We perform this analysis on all SDSS
spectra, in preference to using the SPECTRO1D measurements of these lines, as
these emission lines are the primary diagnostics of on--going star--formation
in a galaxy and thus, we are interested in detecting any evidence of these
lines in our HDS galaxies. As discussed in Section \ref{Hdelta}, the
flux--summing technique is better for the lower signal--to--noise spectra,
while the Gaussian--fitting method of SPECTRO1D is optimal for higher
signal--to--noise detections of these emission lines, especially in the case
of \ha\, where SPECTRO1D de--blends the \ha\, and \nii\, lines.

We use the same flux--summing methodology as discussed above for the \hd\,
line, except we only use one wavelength window centered on the two emission
lines.  We list in Table \ref{tab:wavelength} the wavelength intervals used in
summing the flux for the \oii\, and \ha\, emission lines and the continuum
regions around these lines. Once again, the continuum flux per pixel for each
emission line was estimated using linear interpolation of the continuum
estimated either side of the emission lines (weighted by the inverse square of
the errors on the pixel values during a line fitting procedure). We again
iterate the continuum fit once rejecting $3\sigma$ outliers to the original
continuum fit. We do not de--blend the \ha\, and \nii\, lines and as a result,
some of our \ha\, EW measurements may be over--estimated. However the
contamination is less than 5\% from \nii\, line at 6648\AA\ and less than 30\%
from \nii\, line at 6583\AA. We present estimates of the external error on our
measurements of \oii\, and \ha\, in Section \ref{error}.

In Figure \ref{fig:oii_1d_tomo}, we compare our \oii\, equivalent width
measurements to that from SPECTRO1D for all SDSS spectra. In this paper,
positive EWs are absorption lines and negative EWs are emission lines. There
is a good agreement between the two methods for EW(\oii)$>10$\AA, where the
scatter is $\simless 10\%$. However, at lower EWs, the SPECTRO1D measurement
of \oii\, is systematically larger than our flux--summing method which is the
result of SPECTRO1D fitting a broad Gaussian to the noise in the spectrum.
Furthermore, our continuum measurements for the \oii\, line are estimated
close to the \oii\, line, in a region of the spectrum where the continuum is
varying rapidly with wavelength ({\it e.g.}, D4000 break). Once again, we are
only concerned with making a robust detection of any \oii\, emission, rather
than trying to accurately quantify the properties of the emission line.
Therefore, we prefer our non--parametric method, especially for the low
signal--to--noise cases.

In Figure \ref{fig:ha_1d_tomo}, we compare our \ha\, equivalent width
measurements against that of SPECTRO1D for all SDSS spectra regardless of
their \hd\ EW. The two locii of points seen in this figure are caused by
contamination in our estimates of \ha\, due to strong emission lines in AGNs,
{\it i.e.}, the top locus of points have larger EWs in our flux--summing
method than measured by SPECTRO1D due to contamination by the \nii\,
lines. This is confirmed by the fact that the top locus of points is dominated
by AGNs. At low \ha\, EWs, we again see a systematic difference between our
measurements and those of SPECTRO1D, with SPECTRO1D again over--estimating the
\ha\, line because it is jointly fitting multiple Gaussians to low
signal--to--noise detections of the \ha\, and \nii\, emission lines. Finally,
we do not make any correction for extinction and stellar absorption on our
flux--summed measurements of \ha (see Section \ref{emfilling}).

\subsection{Emission--Filling of the \hd\, Line}\label{emfilling}

As mentioned above, our measurements of the \hd\, absorption line can be
affected by emission--filling, {\it i.e.}, \hd\, emission at the bottom of
the \hd\, absorption line. This problem could be solved by fitting two
Gaussians to the \hd\, line; one for absorption, one for emission. We found
however that this method is only reliable for spectra with a signal--to--noise
of $>20$ and, as shown in Figure \ref{fig:sng}, this is only viable for a
small fraction of our spectra. Therefore, we must explore an alternative
approach for correcting for this potential systematic bias; however, we stress
that the sense of any systematic bias on our non--parametric summing method
would be to always decrease (less absorption) the observed EW of the \hd\
absorption line and thus our technique gives a lower limit to the amount of
\hd\, absorption in the spectrum.

To help rectify the problem of emission--filling, we have used the \ha\, and
\hb\ emission lines (where available) to jointly constrain the amount of
emission--filling at the \hd\, line as well as estimate the effects of
internal dust extinction in the galaxy. Furthermore, our estimates of the
emission--filling are complicated by the effects of stellar absorption on the
\ha\, and \hb\ emission lines. In this analysis, we have used the SPECTRO1D
measurements of \ha\, and \hb\ lines in preference to our flux--summing
technique discussed in Section \ref{oiiha}, because the emission--filling
correction is only important in strongly star--forming galaxies where the
\ha\, and \hb\ emission lines are well fit by a Gaussian and, for the \ha\,
line, require careful de--blending from the \nii\, lines.

To solve the problem of emission--filling, we have adopted two different
methodologies which we describe in detail below.  The first method is an
iterative procedure that begins with a initial estimate for the amount of
stellar absorption at the \hb\ and \ha\, emission lines, {\it i.e.}, we assume
EW (absorption) = 1.5\AA\ and H$\alpha$ EW (absorption) = 1.9\AA\ (see
Poggianti \& Barbaro 1997; Miller \& Owen 2002). Then, using the observed
ratio of the \ha\, and \hb\ emission lines (corrected for stellar absorption), in
conjunction with an attenuation law of $\tau=A\,\lambda^{-0.7}$ (Charlot \&
Fall 2000) for galactic extinction and a theoretical \ha\, to \hb\ ratio of
2.87 (case B recombination; Osterbrock 1989), we solve for $A$ in the
attenuation law and thus gain extinction--corrected values for both the \hb\
and \ha\, emission lines.  Next, using the theoretical ratio of \hb\ emission
to \hd\, emission, we obtain an estimate for the amount of emission--filling
(extinction--corrected) in the \hd\, absorption line.  We then correct the
observed \hd\, absorption EW for this emission--filling and, assuming the
EW(H$\delta$) absorption is equal to EW(\hb) absorption and EW(\ha) absorption
is equal to 1.3 + $0.4\times$ EW(\hb) absorption (Keel 1983), we obtain new estimates for
the stellar absorption at the \ha\, and \hb\ emission lines, {\it i.e.}, where
we begun the iteration.  We iterate this calculation five times, but on
average, a stable solution converges after only one iteration. 

Our second method uses the D4000 break to estimate the amount of
stellar absorption at H$\beta$, using EW(H$\beta$) $= -5.5\times\,{\rm
D4000} + 11.6$ (Poggianti \& Barbaro 1997; Miller \& Owen 2002). Then,
assuming EW(\ha) absorption is equal to 1.3 + $0.4\times$
EW(H$\beta$) absorption, we obtain an measurement for the amount of
stellar absorption at both the \ha\, and H$\beta$ absorption lines.  As
in the first method above, we use the Charlot \& Fall (2000)
attenuation law, and the theoretical \ha\, to H$\beta$ ratio, to solve
for the amount of extinction at \ha\, and H$\beta$, and then use these
extinction--corrected emission lines to estimate the amount of
emission--filling at \hd. We do not iterate this method, as we have
used the measured D4000 break to independently estimate the amount
of stellar absorption at \hb\ and \ha.
 
\begin{figure}
\plotone{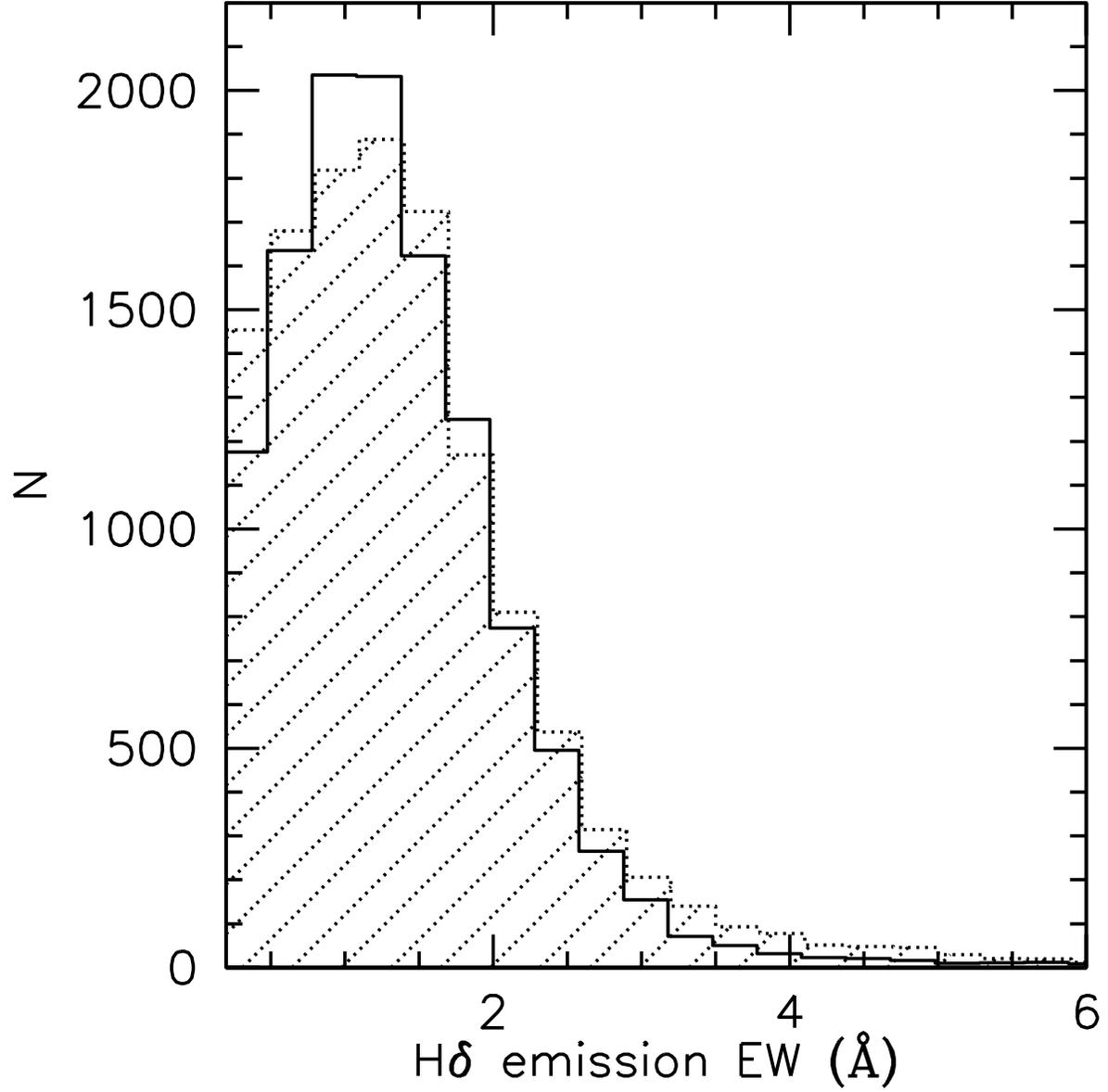}
\caption{
\label{fig:hd_emission_hist}
The amount of emission filling correction of H$\delta$ EW. A solid line is
 for the iteration method (EF1) and the shaded histogram uses the D4000 method (EF2).
}
\end{figure}

\begin{figure}
\centering{\includegraphics[scale=0.3]{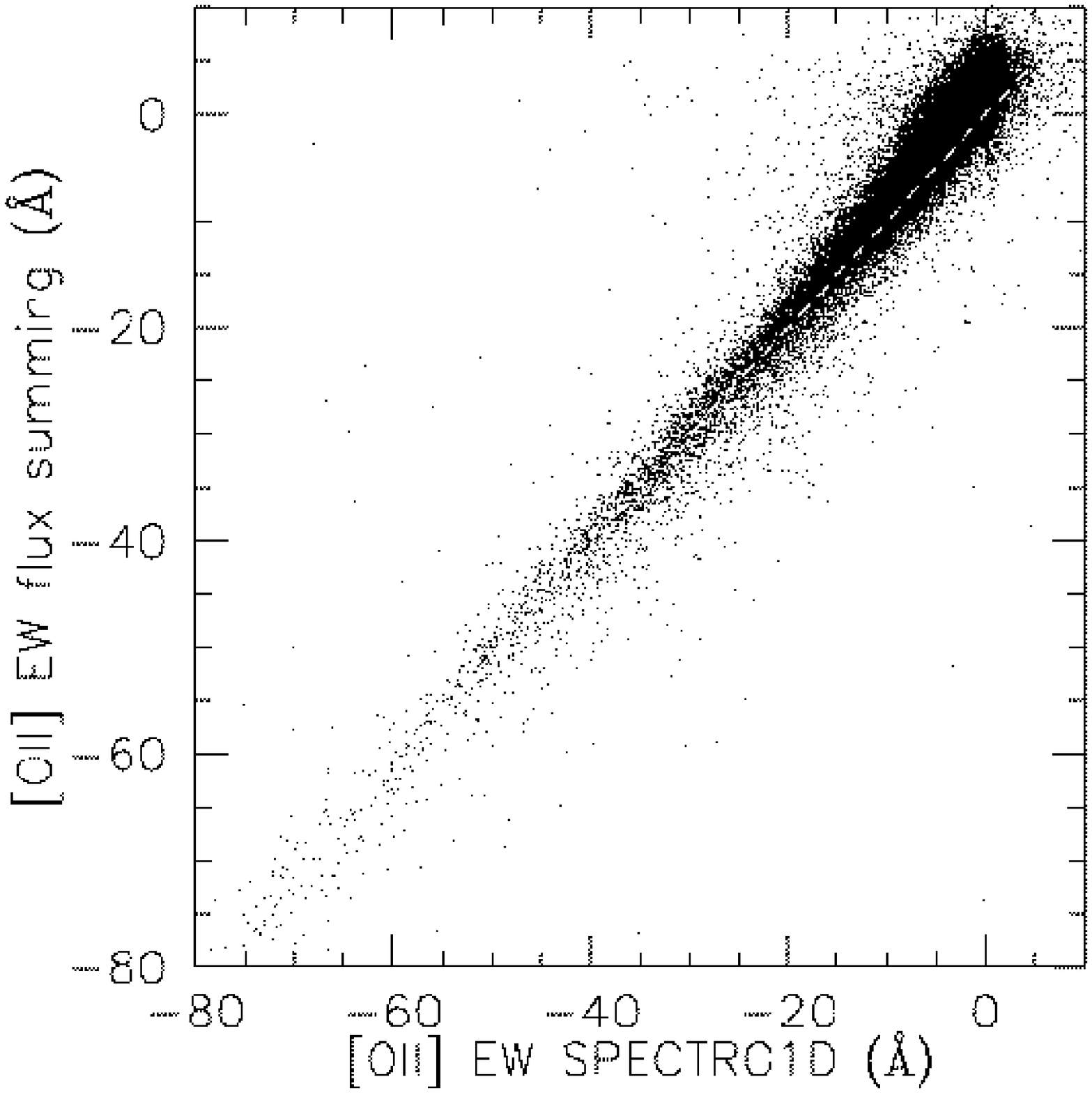}
\includegraphics[scale=0.3]{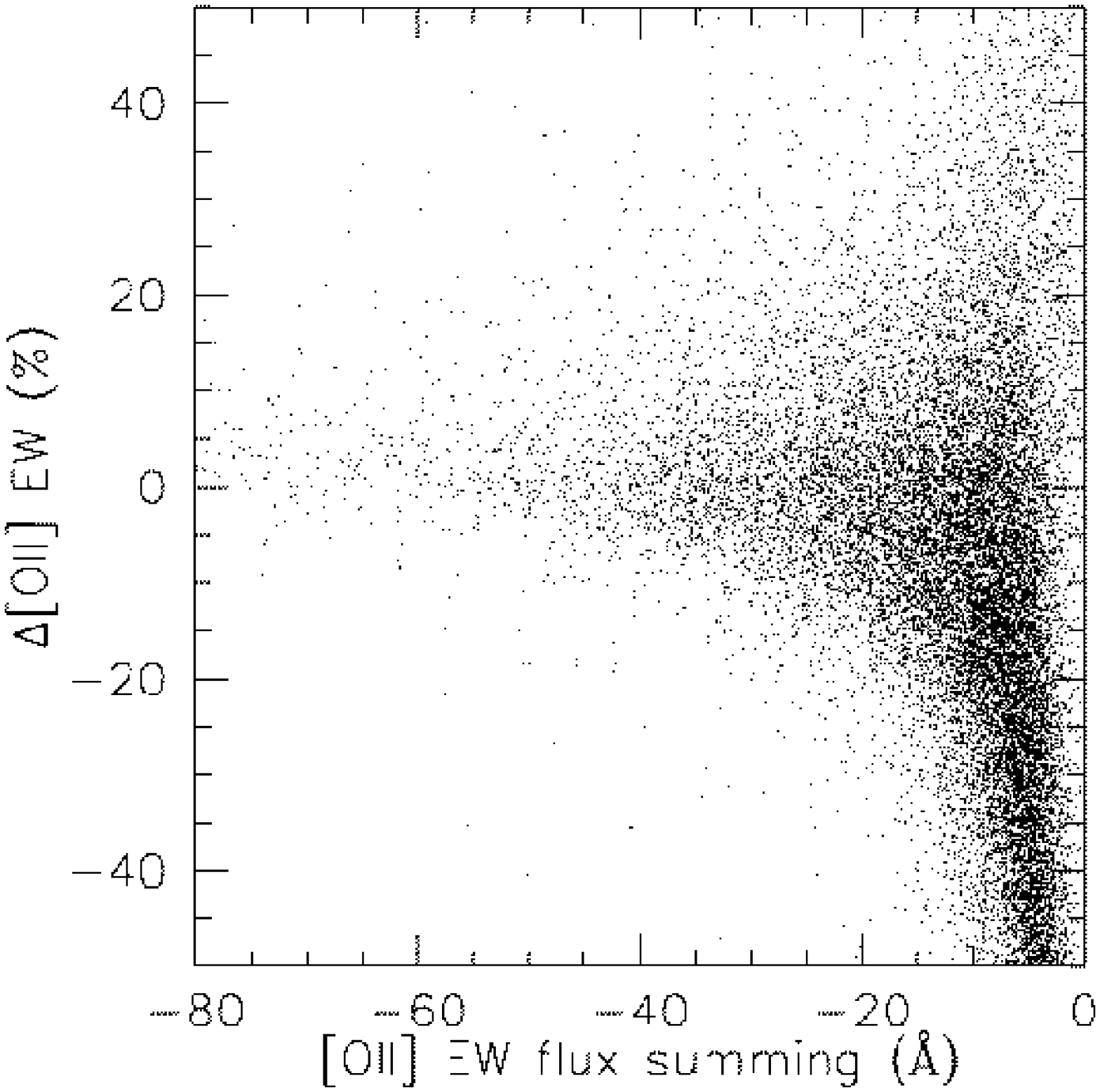}}
\caption{
\label{fig:oii_1d_tomo}
In the left panel, we present the comparison of our \oii\ EW measurements (flux
summing) and those of SPECTRO1D (Gaussian fitting) for all SDSS spectra regardless of their \hd\ EWs. In the right panel, we
plot the percentage difference between these two measurements. Positive
percentages mean our flux summing method has a larger value.  }
\end{figure}

\begin{figure}
\centering{\includegraphics[scale=0.3]{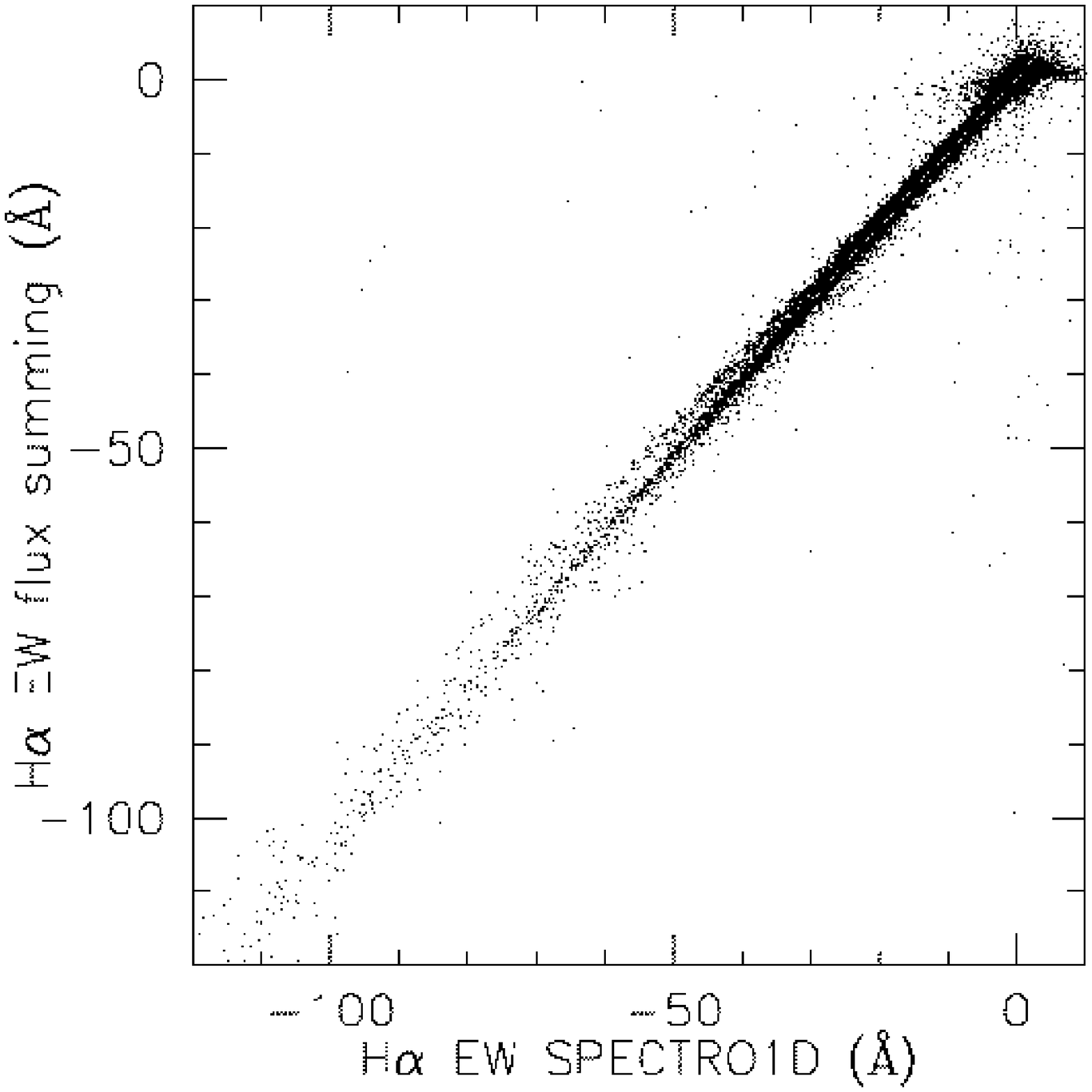}
\includegraphics[scale=0.3]{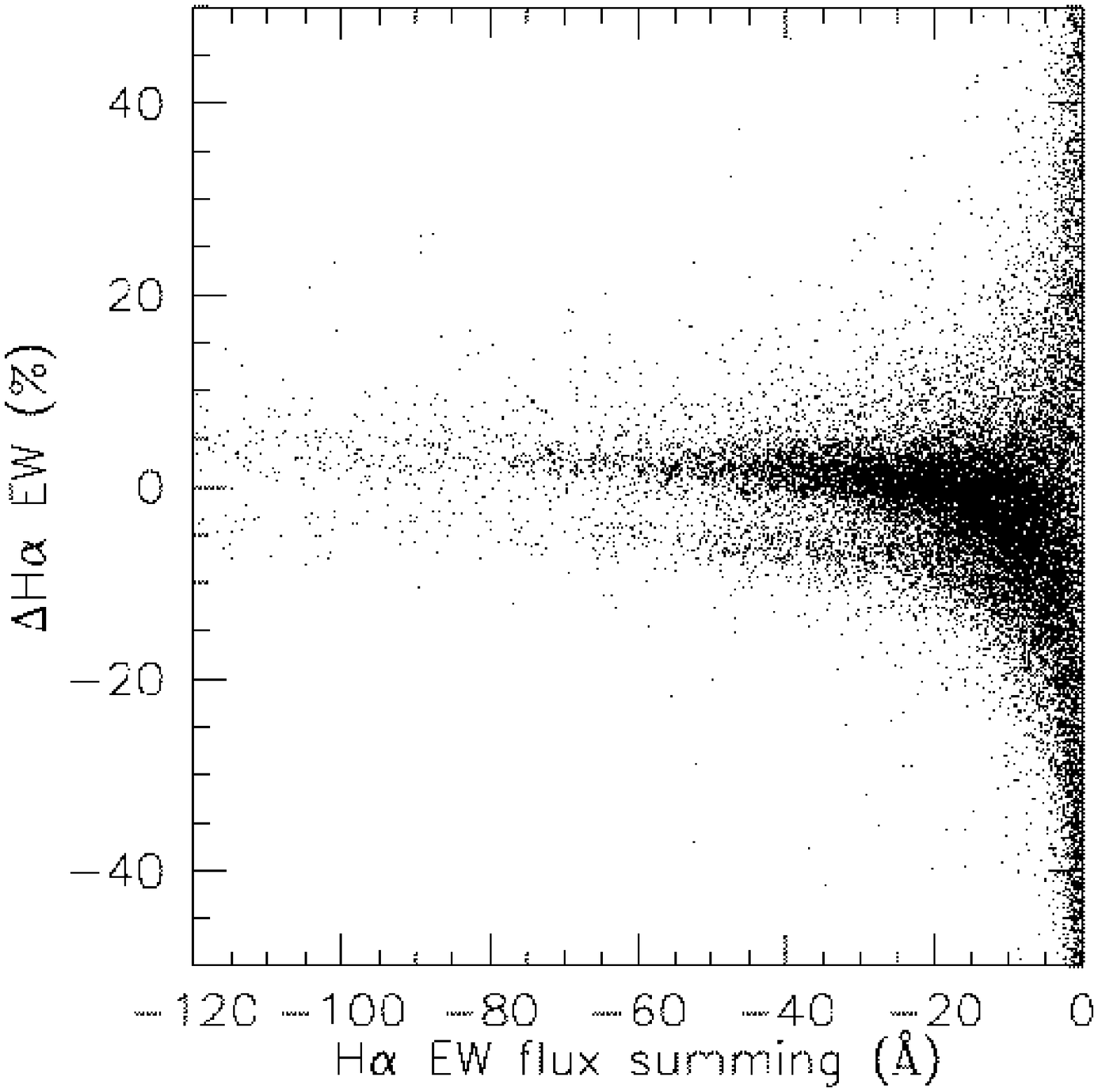}}
\caption{
\label{fig:ha_1d_tomo}
In the left panel, we present the comparison of our \ha\ EW measurements (flux
summing) and those of SPECTRO1D (Gaussian fitting) for all SDSS spectra regardless of their \hd\ EWs. In the right panel, we
plot the percentage difference between these two measurements. Positive
percentages mean our flux summing method has a larger value.  }
\end{figure}

We have applied these two methods to all our SDSS spectra, except for any
galaxy that possesses a robust detection of an Active Galactic Nucleus (AGN)
based on the line indices discussed in Kewley et al. (2002) and Gomez et
al. (2003). For these AGN classifications, we have used the SPECTRO1D emission
line measurements. We also stop our emission--filling correction if the ratio
of the \hb\ and \ha\, line becomes unphysical, {\it i.e.}, greater than
2.87. By definition, the emission--filling correction increases our observed
values of the \hd\, absorption line, with a median correction of 15\% in the
flux of the \hd\, absorption line. In Figure \ref{fig:hd_emission_hist}, we
show the distributions of H$\delta$ emission EWs calculated using the two
methods described above in a solid (iteration) and a dashed (D4000) line,
respectively. It is re--assuring that overall these two methods give the same
answer and have similar distributions.

\subsection{External Errors on our Measured Equivalent Widths}\label{error}

\begin{figure}
\plotone{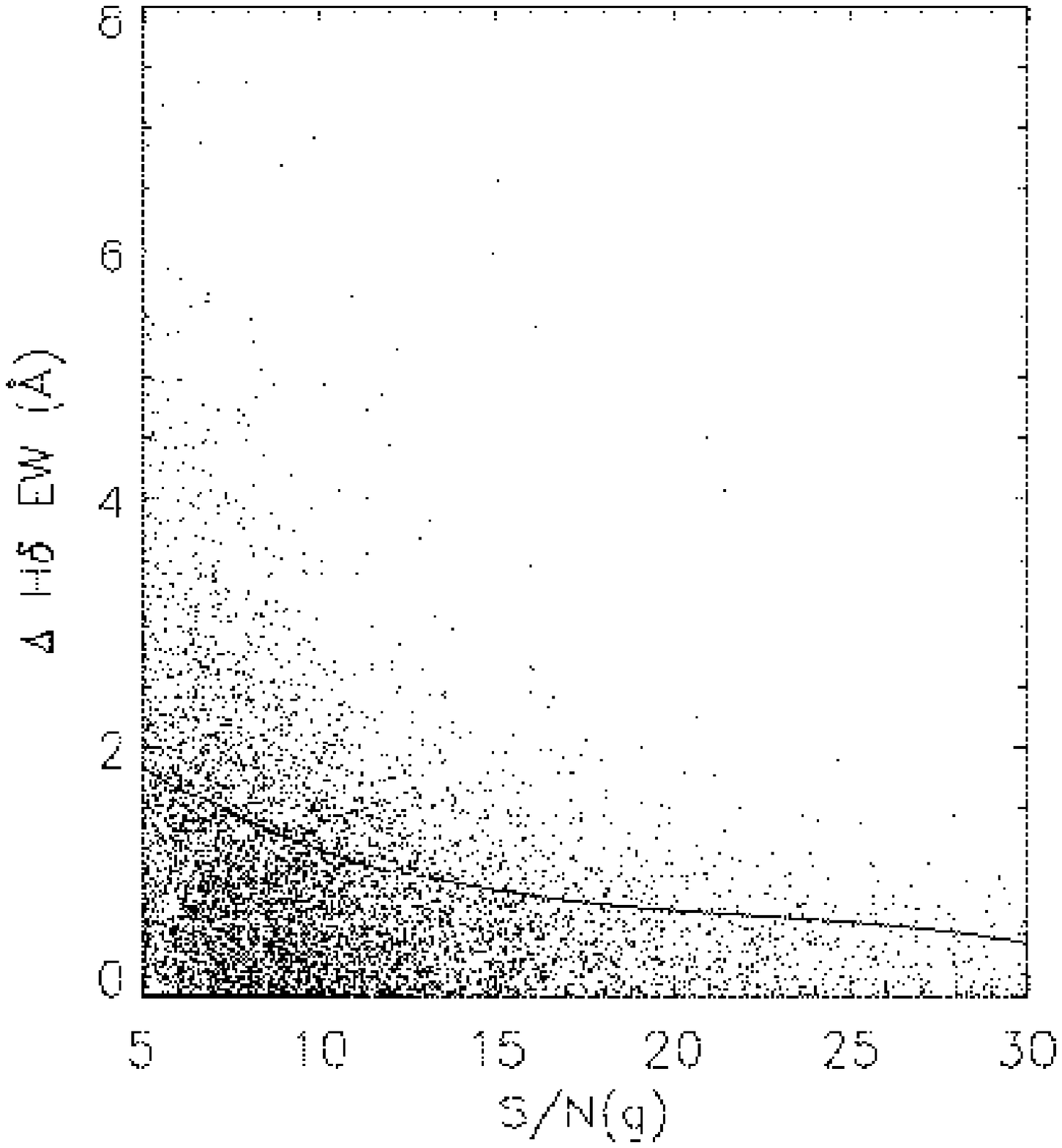}
\caption{
\label{fig:err_from_double_obs_hd}
The absolute difference in the measured \hd\, EW (\AA) for duplicate
observations of SDSS galaxies as a function of signal--to--noise (the
lower of the two signal--to--noise ratios has been plotted here). The
solid line shows the $1\sigma$ polynomial line fitted to the
distribution of errors (as a function of signal--to--noise).  }
\end{figure}

\begin{figure}
\plotone{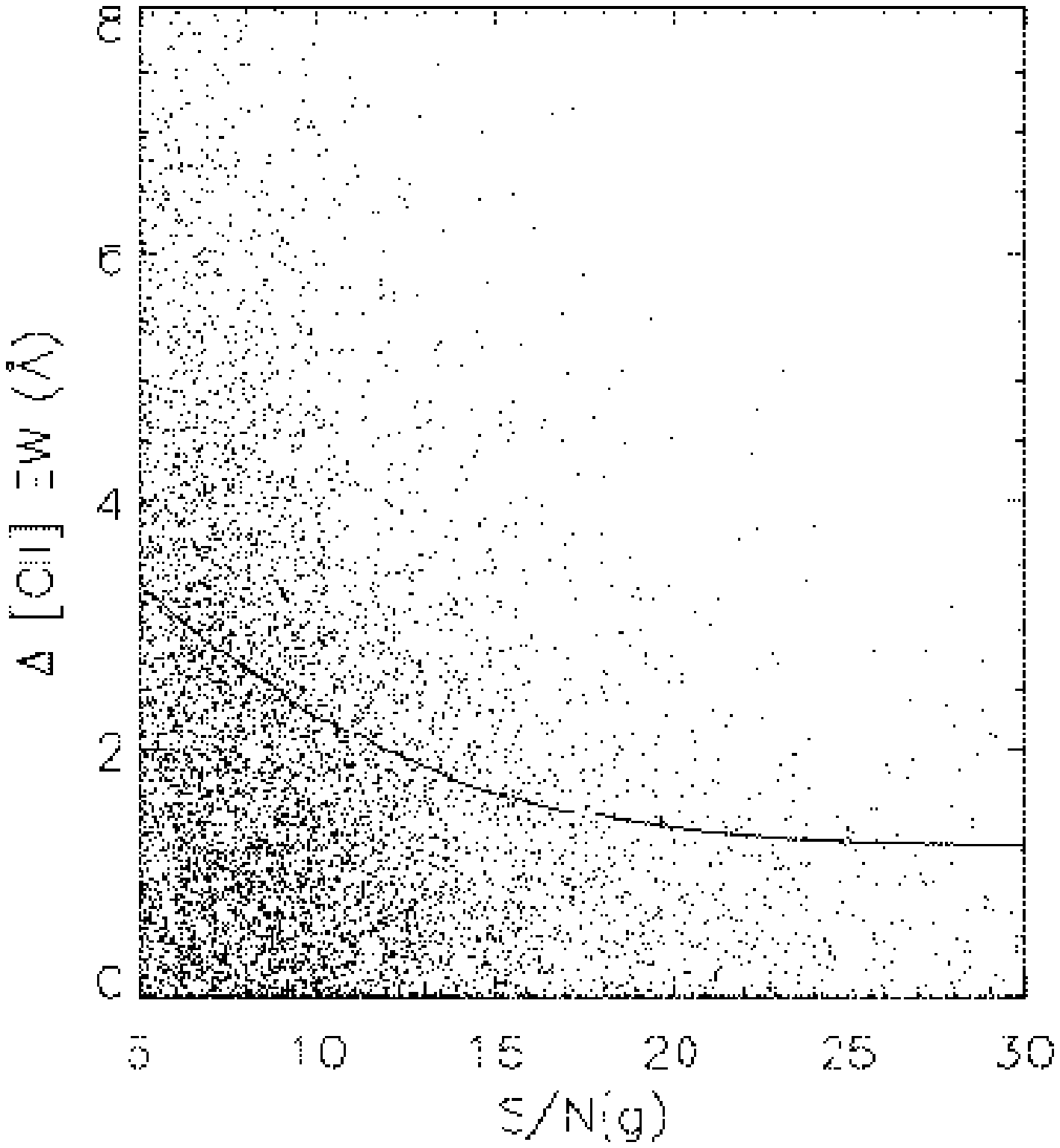}
\caption{
\label{fig:err_from_double_obs_oii}
The absolute difference in the \oii\, EW (\AA) for duplicate
observations of SDSS galaxies as a function of signal--to--noise (the
lower of the two signal--to--noise ratios has been plotted here). The
solid line shows the $1\sigma$ polynomial line fitted to the
distribution of errors (as a function of signal--to--noise).}
\end{figure}

\begin{figure}
\plotone{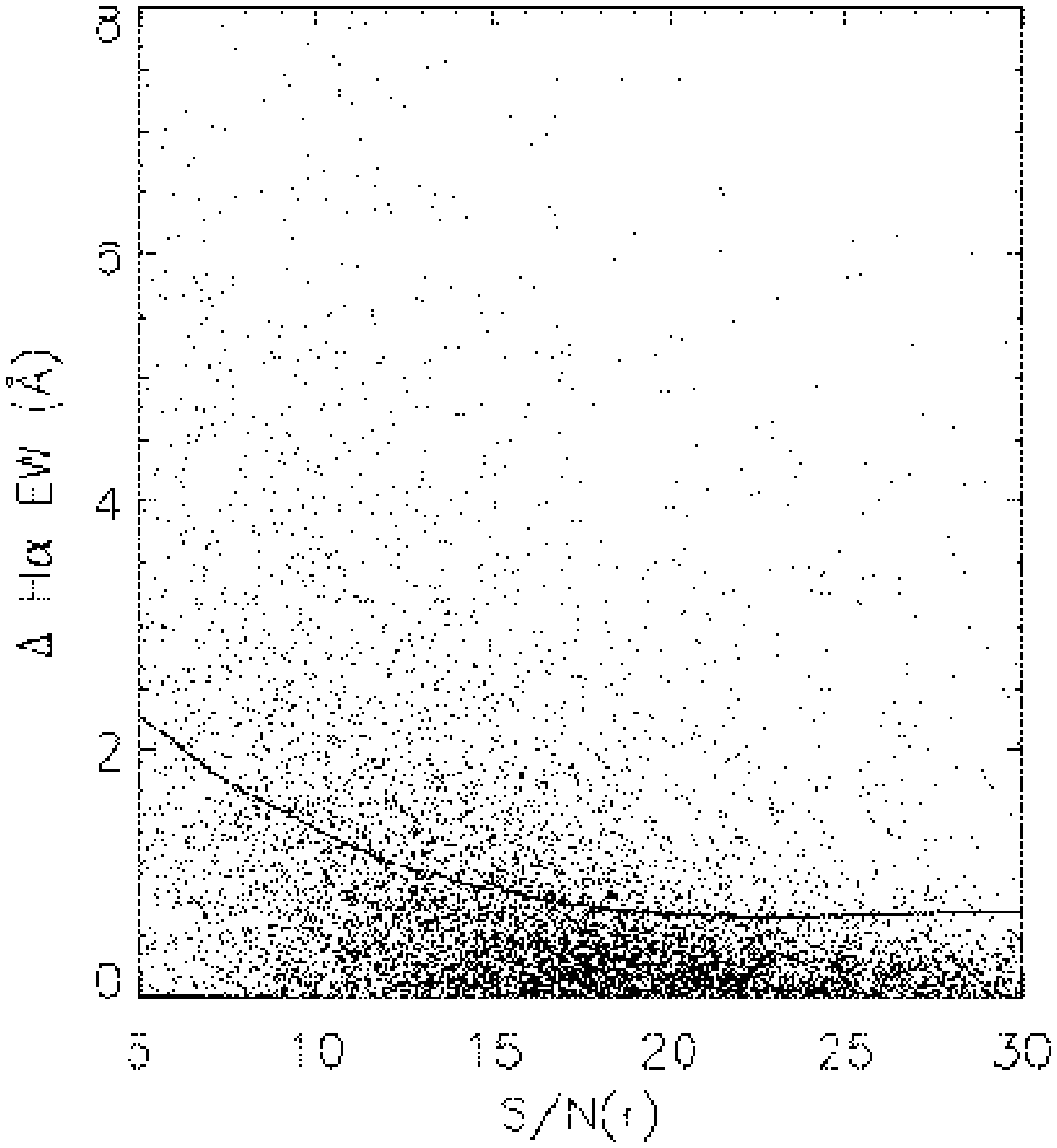}
\caption{
\label{fig:err_from_double_obs_ha}
The absolute difference in the \ha\, EW (\AA) for duplicate observations
of SDSS galaxies as a function of signal--to--noise (the lower of the
two signal--to--noise ratios has been plotted here).  The solid line
shows the $1\sigma$ polynomial line fitted to the distribution of
errors (as a function of signal--to--noise).  }
\end{figure}

\begin{figure}
\centering{\includegraphics[scale=0.4]{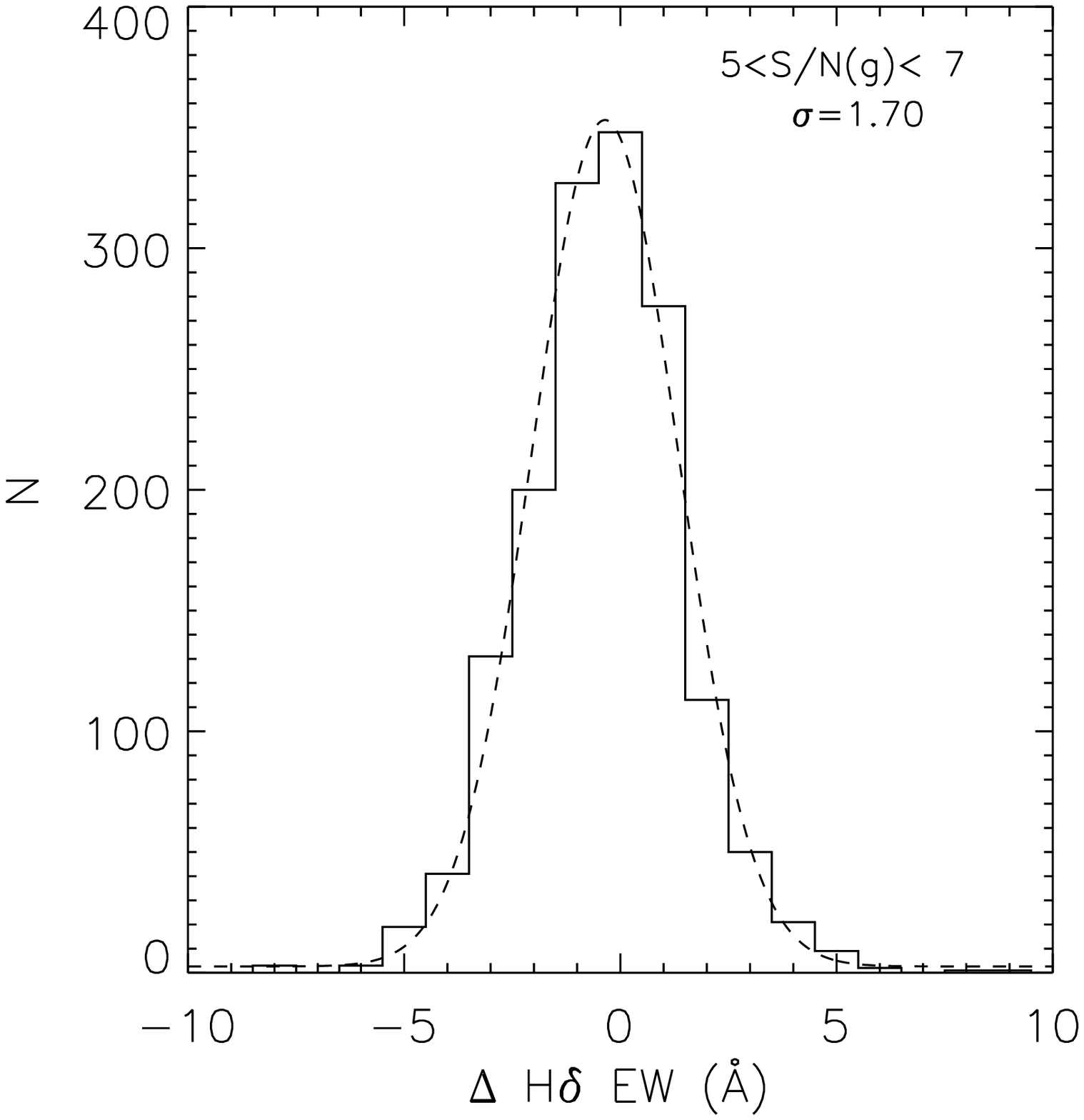}
\includegraphics[scale=0.4]{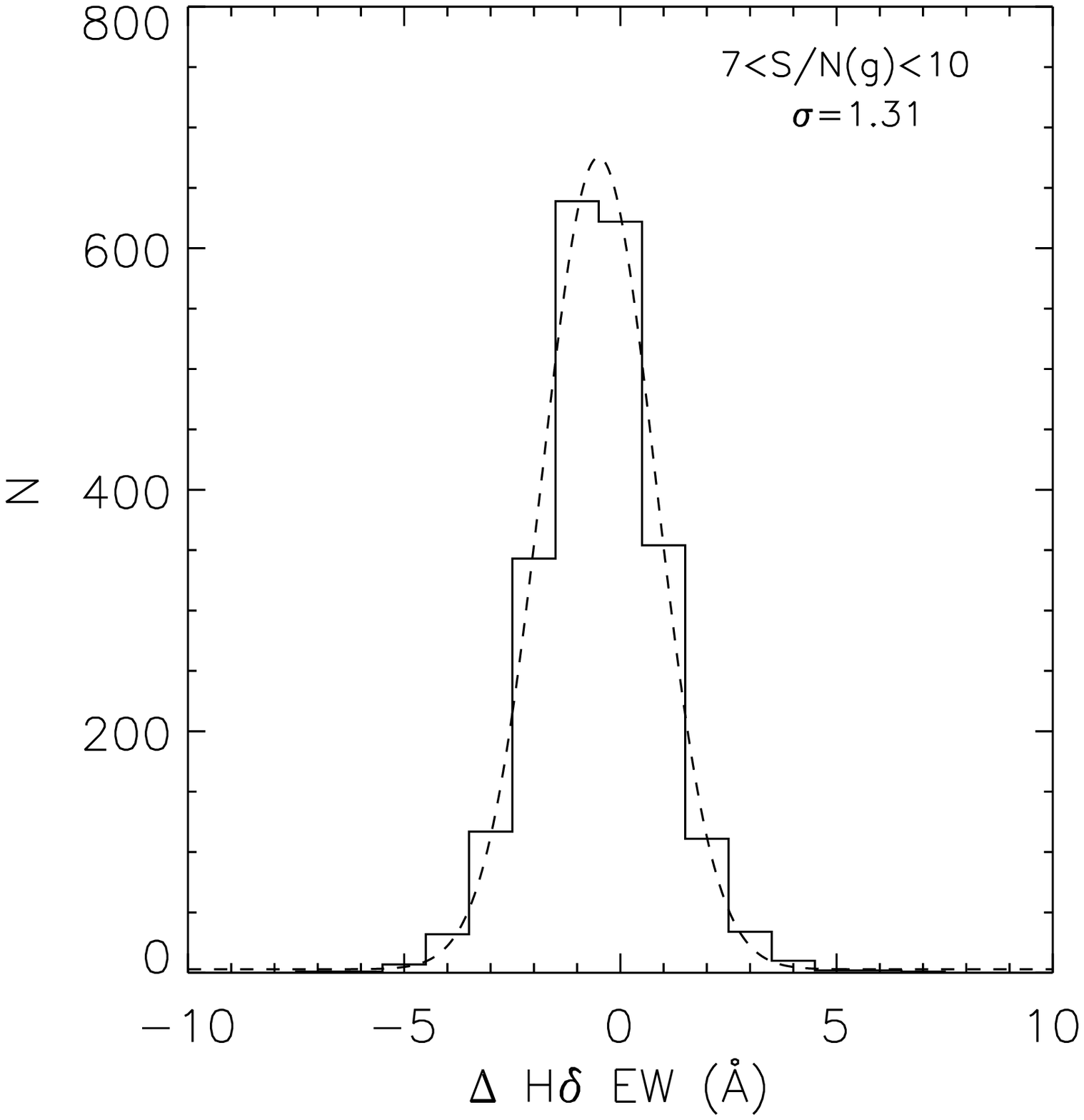}}
\centering{\includegraphics[scale=0.4]{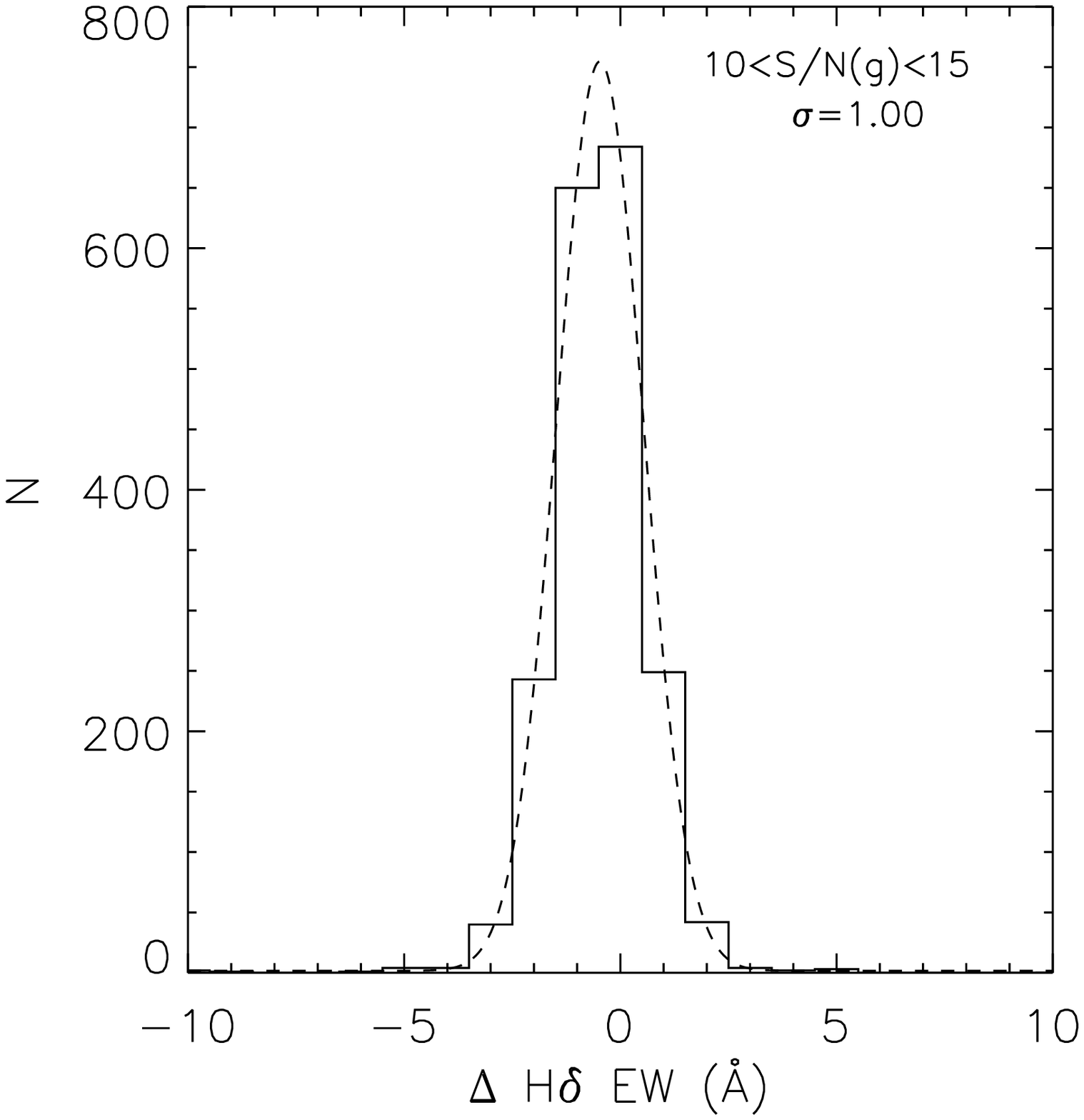}
\includegraphics[scale=0.4]{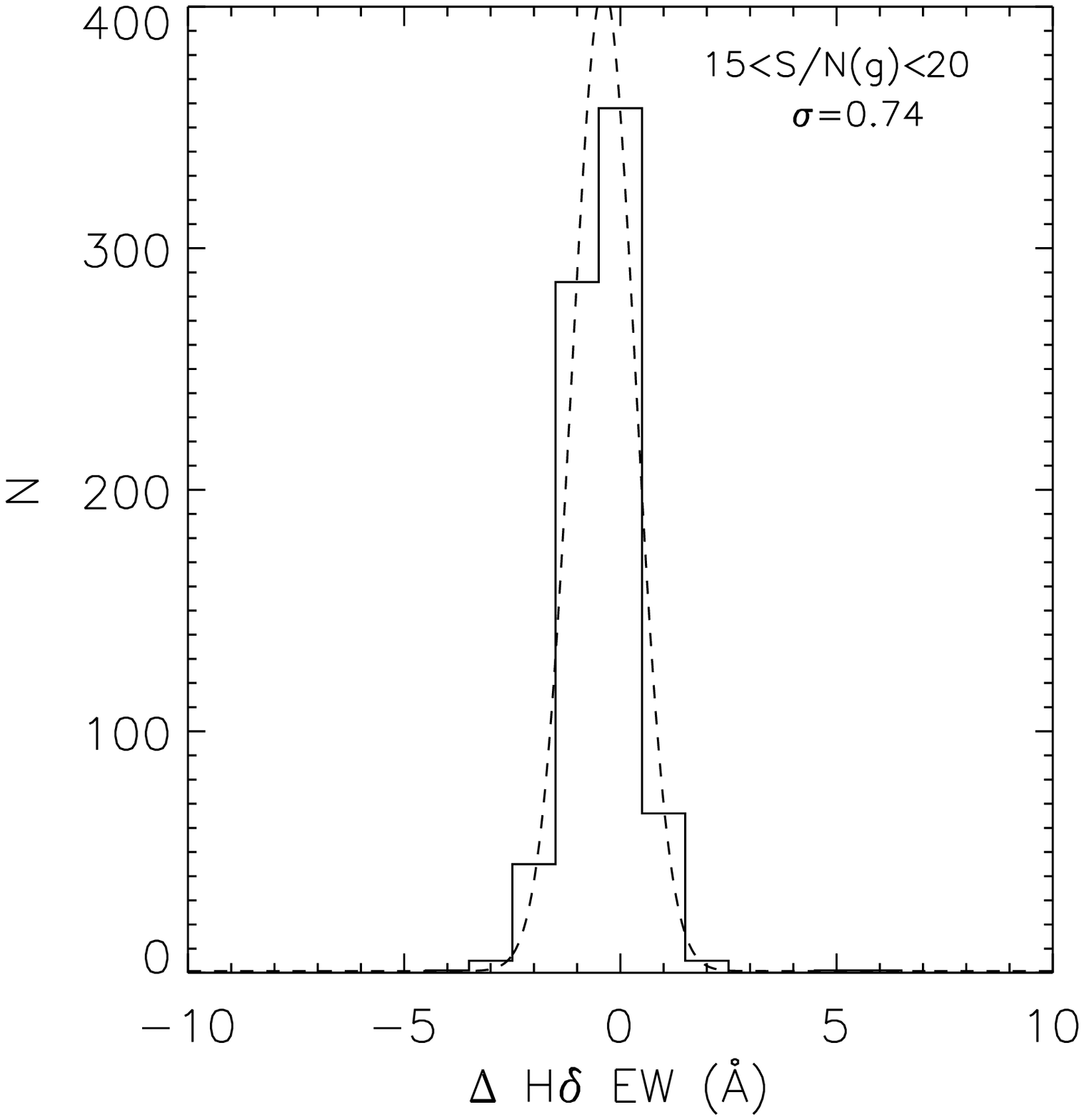}}
\caption{
\label{fig:err_from_double_obs_hd_gauss_fit}
We present the distribution of differences for the \hd\, line from our
duplicate observations. The four panels denote four different bins in
signal--to--noise, {\it i.e.}, clockwise from the top--left panel, we have s/n
$<7$, $7<{\rm s/n} <10$, $15<{\rm s/n} <20$ and $10<{\rm s/n} <15$.  We show
as a dotted line the best fit gaussian to these distributions, which was then
used to determine the $1\sigma$ error (shown for each panel) on \hd\, EW as a
function of signal--to--noise.}
\end{figure}

\begin{figure}
\centering{\includegraphics[scale=0.4]{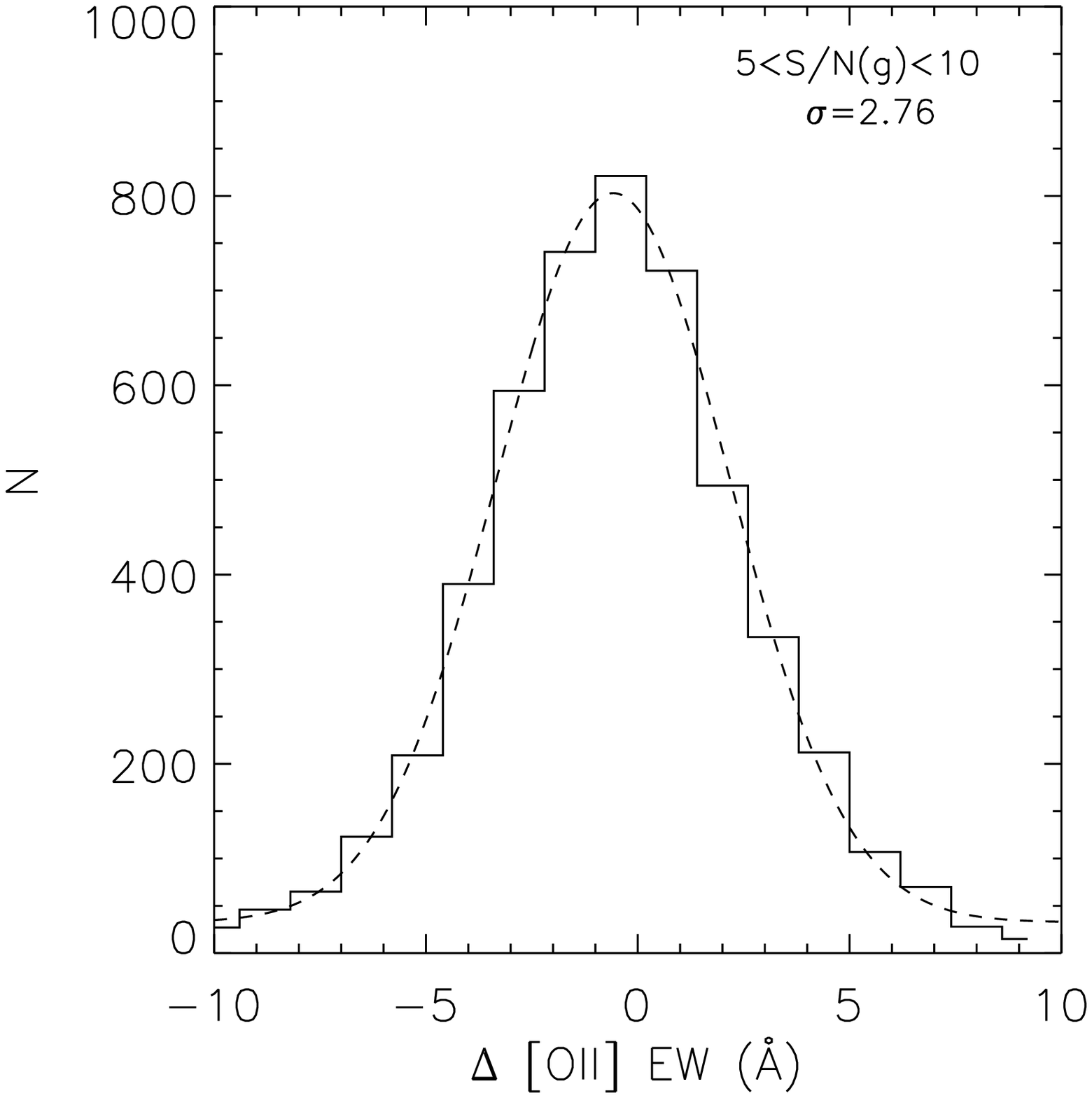}
\includegraphics[scale=0.4]{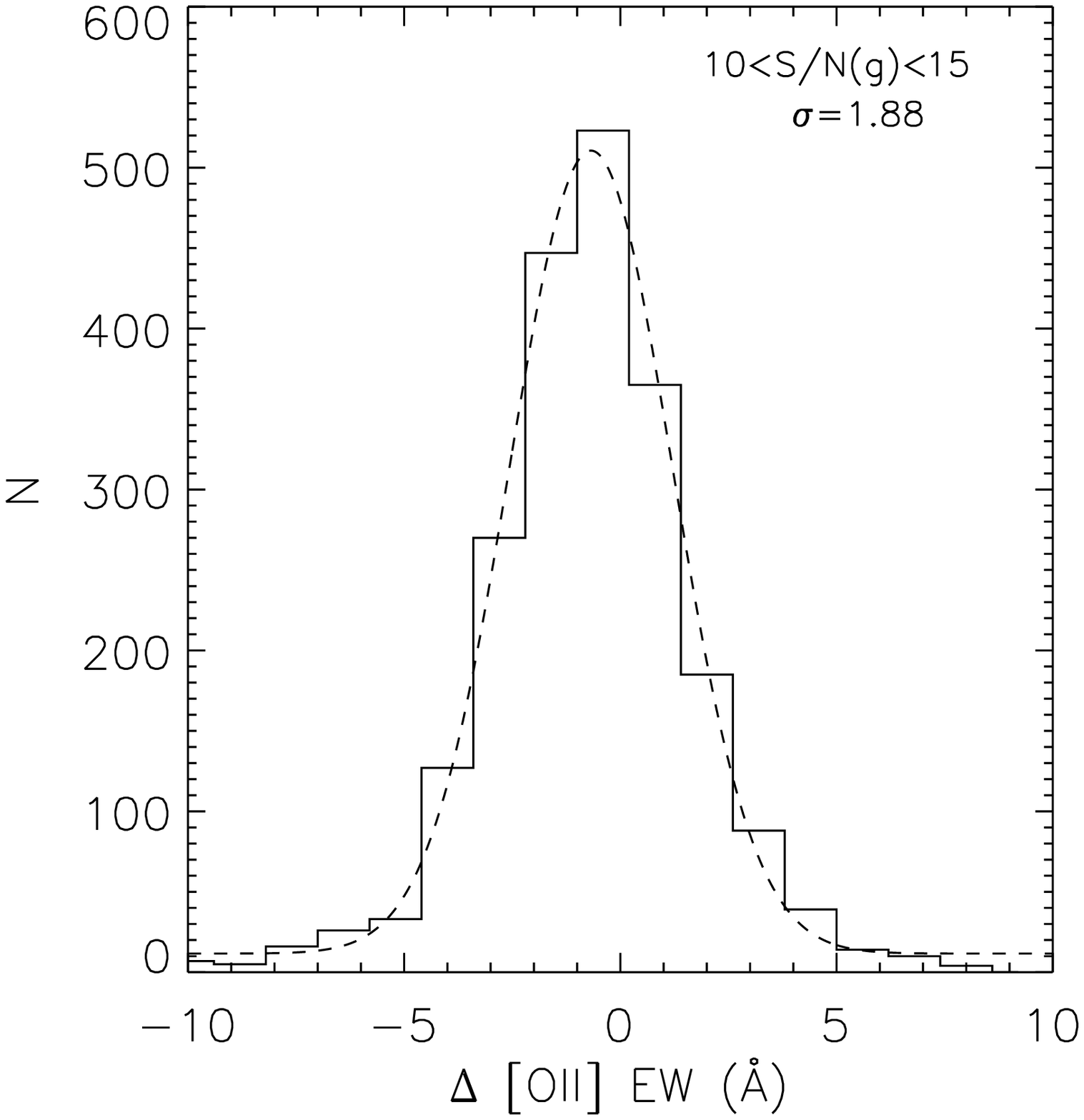}}
\centering{\includegraphics[scale=0.4]{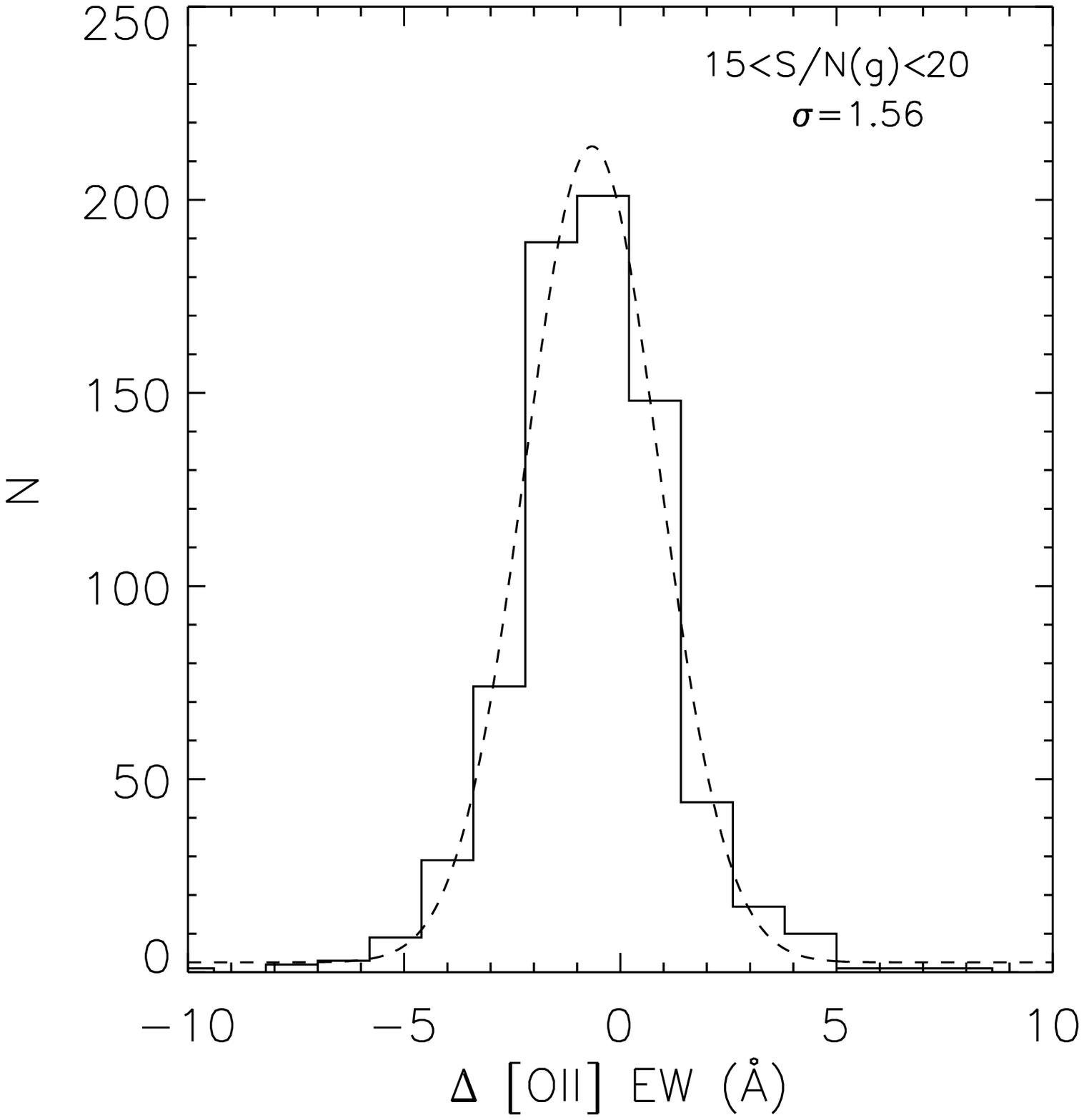}
\includegraphics[scale=0.4]{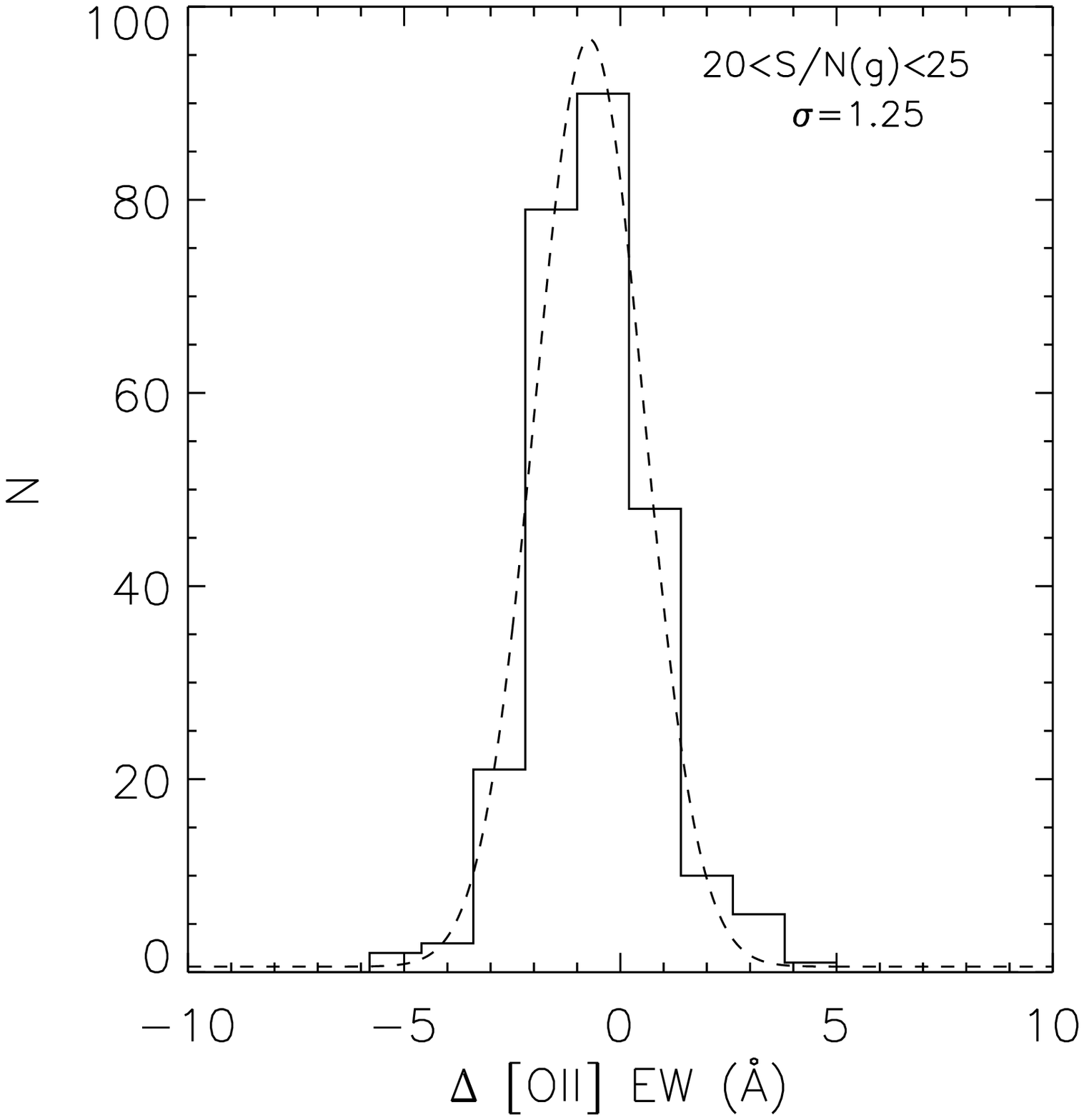}}
\caption{
\label{fig:err_from_double_obs_hd_gauss_fit_oii}
We present the distribution of differences for the \oii\, line
from our duplicate observations. The four panels denote four different
bins in signal--to--noise, {\it i.e.}, clockwise from the top--left
panel, we have $5<{\rm s/n} <10$, $10<{\rm s/n} <15$, $20<{\rm s/n}
<25$ and $15<{\rm s/n} <20$.  We show as a dotted line the best fit
gaussian to these distributions, which was then used to determine the
$1\sigma$ error on \oii\, EW as a function of signal--to--noise.
}\end{figure}

\begin{figure}
\centering{\includegraphics[scale=0.4]{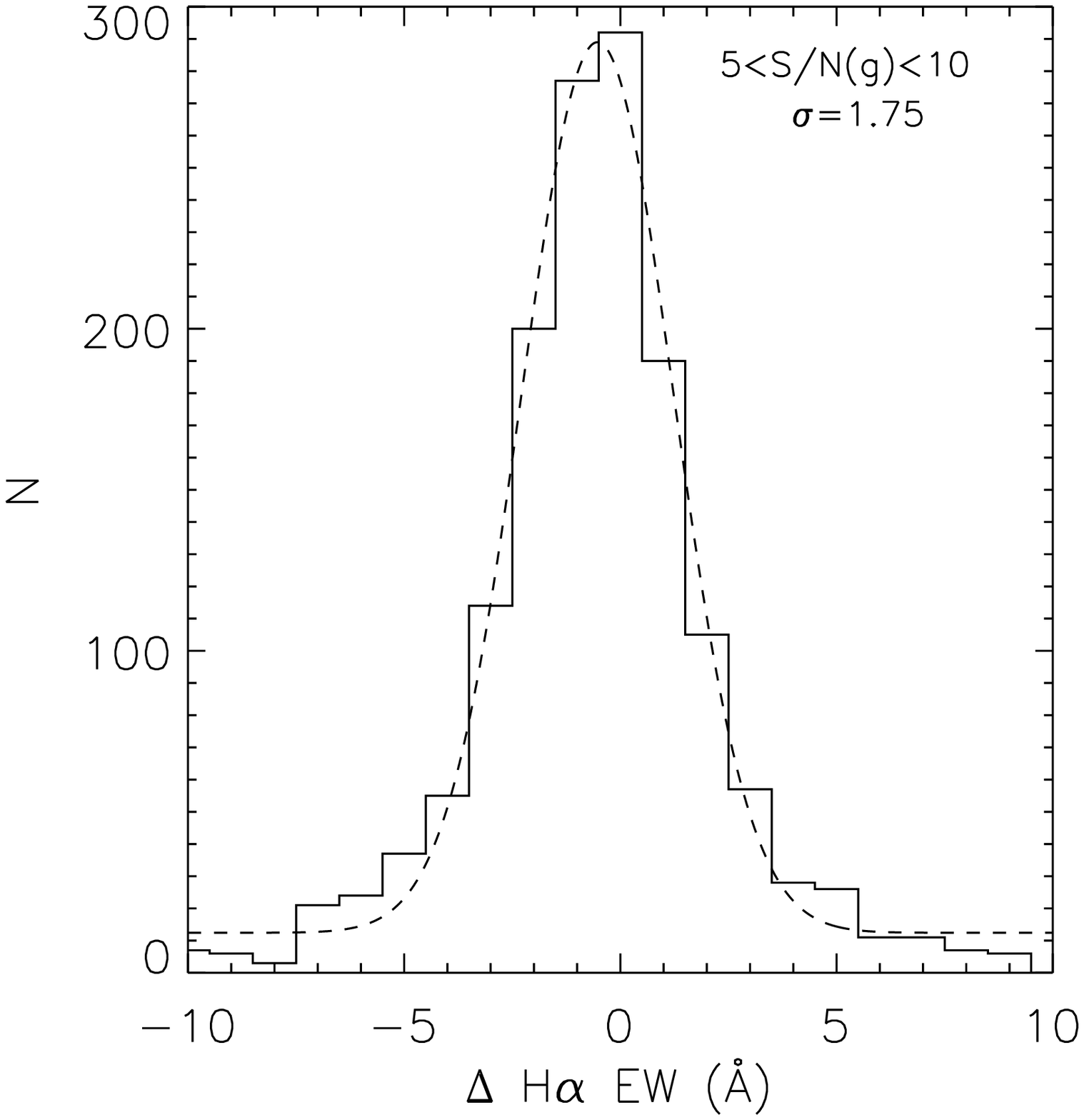}
\includegraphics[scale=0.4]{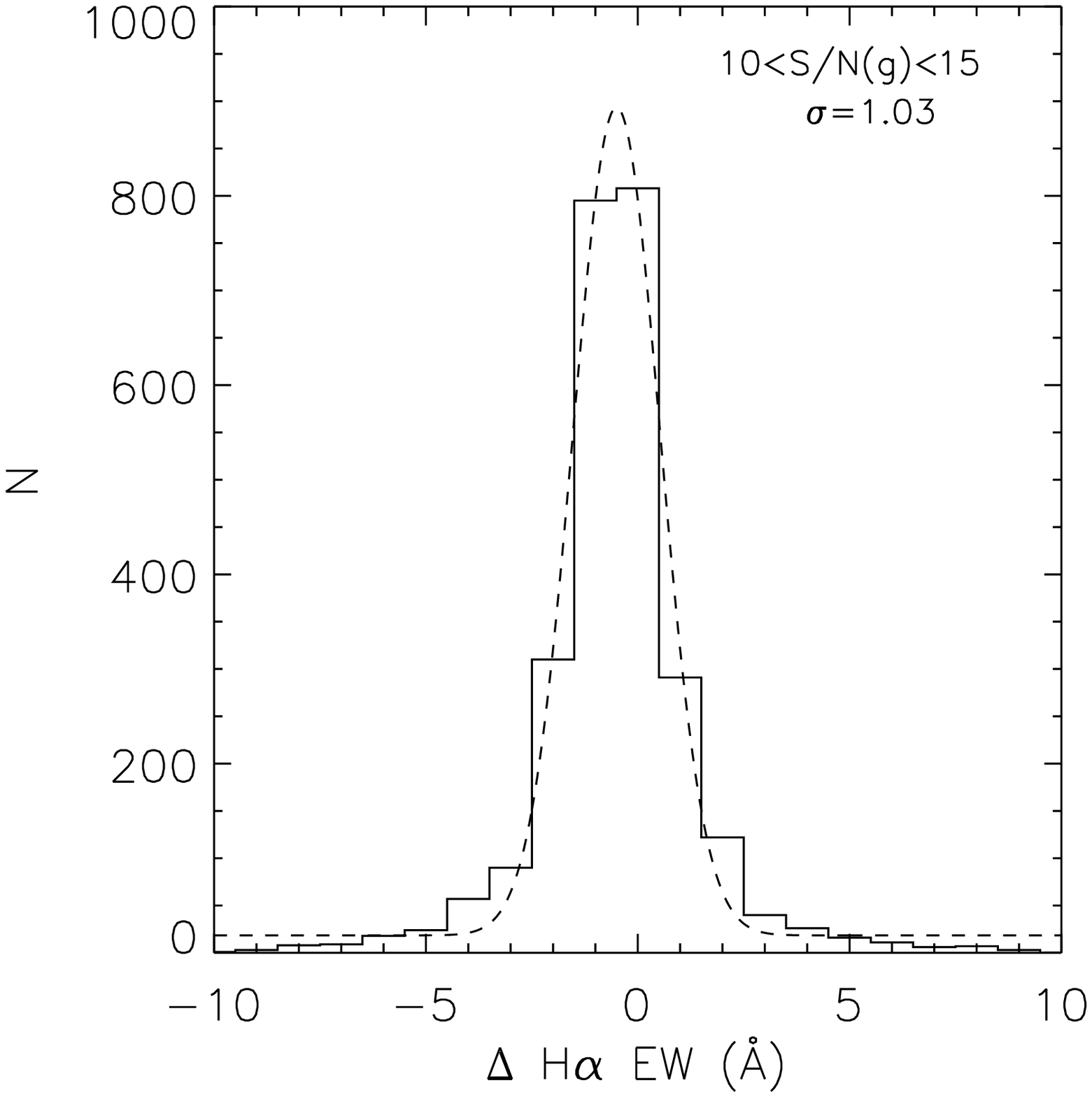}}
\centering{\includegraphics[scale=0.4]{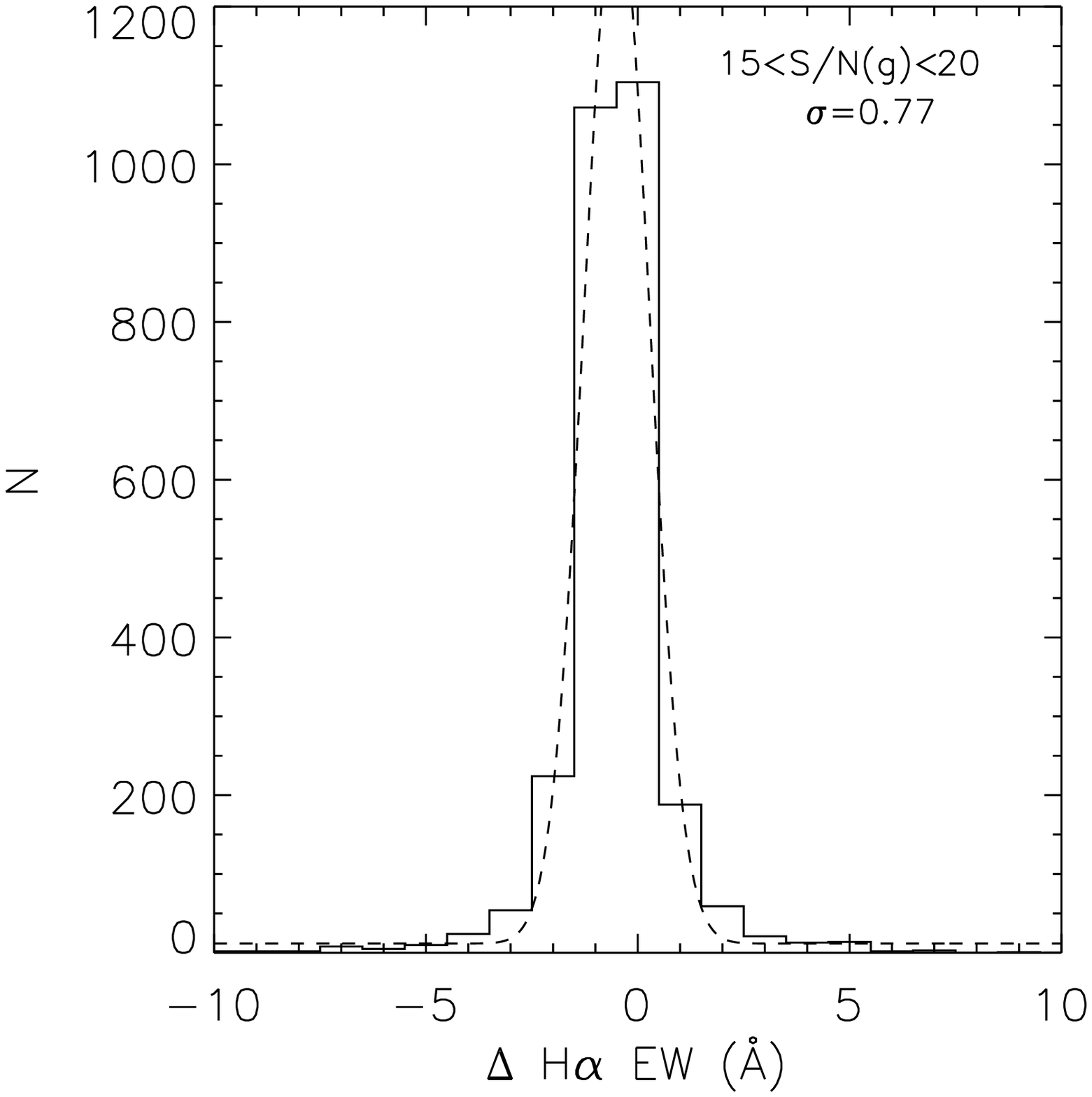}
\includegraphics[scale=0.4]{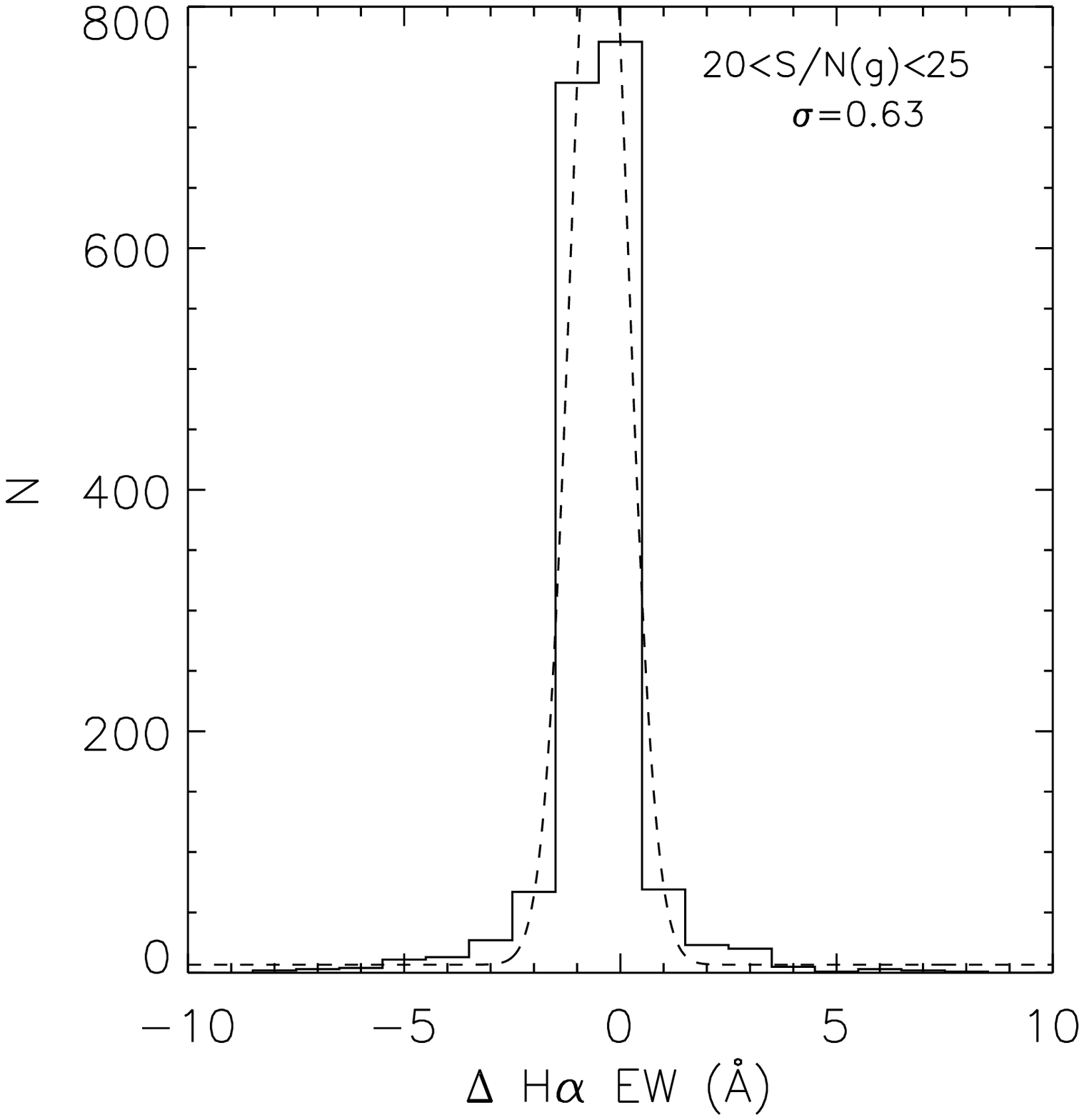}}
\caption{
\label{fig:err_from_double_obs_hd_gauss_fit_ha}
We present the distribution of differences for the \ha\, line from our
duplicate observations. The four panels denote four different bins in
signal--to--noise, {\it i.e.}, clockwise from the top--left panel, we have
$5<{\rm s/n} <10$, $10<{\rm s/n} <15$, $20<{\rm s/n} <25$ and $15<{\rm s/n}
<20$.  We show as a dotted line the best fit gaussian to these distributions,
which was then used to determine the $1\sigma$ error on \ha\, as a function of
signal--to--noise.  }\end{figure}

Before we select our HDS sample of galaxies, it is important to accurately
quantify the errors on our EW measurements.  In our data, there are 11538
galaxies spectroscopically observed twice (see Section \ref{data}), which we
use to quantify the external error on our EW measurements. In Figures
\ref{fig:err_from_double_obs_hd}, \ref{fig:err_from_double_obs_oii} \&
\ref{fig:err_from_double_obs_ha}, we present the absolute difference in
equivalent width of the two independent observations of the \hd\,, \ha\, and
\oii\, lines, as a function of signal--to--noise. In this analysis, we have
used the lower of the two measured signal--to--noise ratios (in SDSS $g$ band
for \hd\, and \hb, or $r$ band for \ha) as any observed difference in the two
measurements of the EW will be dominated by the error in the noisier (lower
signal--to--noise) of the two spectra.  From this data, we determine the
$1\sigma$ error for each line and assign this, as a function of
signal--to--noise, to our EW measurements for each galaxy. We determine the
sigma of the distribution by fitting a Gaussian (as a function of
signal--to--noise) as shown in Figures
\ref{fig:err_from_double_obs_hd_gauss_fit},
\ref{fig:err_from_double_obs_hd_gauss_fit_oii} \&
\ref{fig:err_from_double_obs_hd_gauss_fit_ha}, and then use a 3rd order
polynomial to interpolate between the signal--to--noise bins, thus obtaining
the solid lines shown in Figures \ref{fig:err_from_double_obs_hd},
\ref{fig:err_from_double_obs_oii} \& \ref{fig:err_from_double_obs_ha}.  Using
these polynomial fits, we can estimate the $1\sigma$ error on our EWs for any
signal--to--noise.  The coefficients of the fitted 3rd order polynomial for
each line are given in Table \ref{equation}.

In addition to quantifying the error on \oii, \ha\, and \hd, we have used the
duplicate observations of SDSS galaxies to determine the error on our
emission--filling corrections. Only 400 (564) of the 11538 duplicate
observations of SDSS galaxies have strong \ha\, and H$\beta$ emission lines
which are required for the iterative (D4000) method of correcting for
emission--filling. The errors on the emission--filling correction are only a
weak function of signal--to--noise, so we have chosen to use a constant value
for their error, rather than varying the error as a function of the galaxy
signal--to--noise as done for \hd, \oii\, and \ha\, emission lines. One sigma
errors on the emission correction of \hd\ for the iterative method (EF1)
and the D4000 method (EF2) are 0.57\AA\ and 0.4\AA\ in EW, respectively.

\section{A Catalog of HDS Galaxies}
\label{catalog}

We are now ready to select our sample of HDS galaxies using the non-parametric
measurements of the \hd\, EW ({\it i.e.}, {\rm EW(H$\delta_{\rm max}$)}).  We
begin by imposing the following threshold on EW(H$\delta_{\rm max}$);

\begin{equation}
{\rm EW(H\delta_{\rm max})} - \Delta {\rm EW(H\delta_{\rm max})} > 4\AA,
\label{mainthres}
\end{equation}

\noindent where $\Delta{\rm EW(H\delta_{\rm max})}$ is the $1\sigma$ error on
\hd\, based on the signal--to--noise of the spectrum (see Figure
\ref{fig:err_from_double_obs_hd}). We have chosen this threshold ($4$\AA)
based on visual inspections of the data and our desire to select galaxies
similar to those selected by other authors (Zabludoff et al. 1996; Balogh et
al. 1999; Poggianti et al. 1999), {\it i.e.}, galaxies with strong recent star
formation as defined by the \hd\, line. This threshold (Eqn. \ref{mainthres})
is applied without any emission--filling correction. For the signal--to--noise
ratios of our spectra (Figure \ref{fig:sng}), only galaxies with an observed
\hd\, of $\simgreat 5$\AA\ satisfy Eqn. \ref{mainthres}, which is close to the
$5$\AA\ threshold used by Balogh et al. (1999) to separate normal
star--formating galaxies from post--starburst galaxies (see Figures 8 \& 9 in
their paper). Therefore, our HDS sample should be similar to those already in
the literature, but is still conservative enough to be inclusive of the many
different subsamples of \hd--strong galaxies, like k+a, a+k, A+em and e(a), as
discussed in Pogginati et al. (1999) and Balogh et al. (1999). We will present
a detailed comparison of our HDS sample with models of galaxy evolution in
future papers.

We call the sample of galaxies that satisfy Eqn. \ref{mainthres} ``Sample 1'',
and the sample contains 2526 galaxies from the main SDSS galaxy sample and 234
galaxies targeted for spectroscopy for other reasons, {\it e.g.}, mostly
because they were LRG galaxies (see Eisenstein et al. 2002), or some were
targeted as stars or quasars (see Gordon et al. 2002).

We now apply the emission--filling correction for each galaxy to ${\rm
EW(H\delta_{\rm max})}$ and select an additional sample of galaxies, that were
not already selected in Sample 1 via Eqn \ref{mainthres}, but now satisfy both
the following criteria;

\begin{equation}
\setcounter{equation}{2}
{\rm EW(H\delta_{\rm max})}-\Delta{\rm EW(H\delta_{\rm max})} - \Delta{\rm EW(EF1)}  > 4\AA,\\
\end{equation}

\begin{equation}
\setcounter{equation}{2}
{\rm EW(H\delta_{\rm max})}-\Delta{\rm EW(H\delta_{\rm max})} - \Delta{\rm EW(EF2)}  > 4\AA,
\label{sample2thres}
\end{equation}

\noindent where $\Delta{\rm EW(EF1)}=0.57$\AA\ and $\Delta{\rm
EW(EF2)}=0.4$\AA\ are the $1\sigma$ errors on the iterative method (EF1) and
D4000 method (EF2) of correcting for emission--filling discussed in Section
\ref{error}. Therefore, this additional sample of galaxies represents systems
that would only satisfy the threshold in Eqn \ref{mainthres} because of the
emission--filling correction (in addition to galaxies already selected as part
of Sample 1).  We call this sample of galaxies ``Sample 2'' and it contains
483 galaxies from the main SDSS galaxy sample and 97 galaxies which were again
targeted for spectroscopy for other reasons, {\it e.g.}, LRG galaxies, stars
or quasars. On average, Sample 2 galaxies have strong emission lines because,
by definition, they have the largest emission--filling correction at the \hd\,
line. We have imposed both of these criteria to control the number of extra
galaxies scattered into the sample. If we relaxed these criteria ({\it i.e.},
remove both $\Delta{\rm EW(EF1)}$ and $\Delta{\rm EW(EF2)}$), then the sample
would increase from 483 galaxies (in the main galaxy sample) to 1029. For
completeness, we provide the extra 546 galaxies, which would be included if
these criteria were relaxed, on our webpage.

\begin{figure}
\plotone{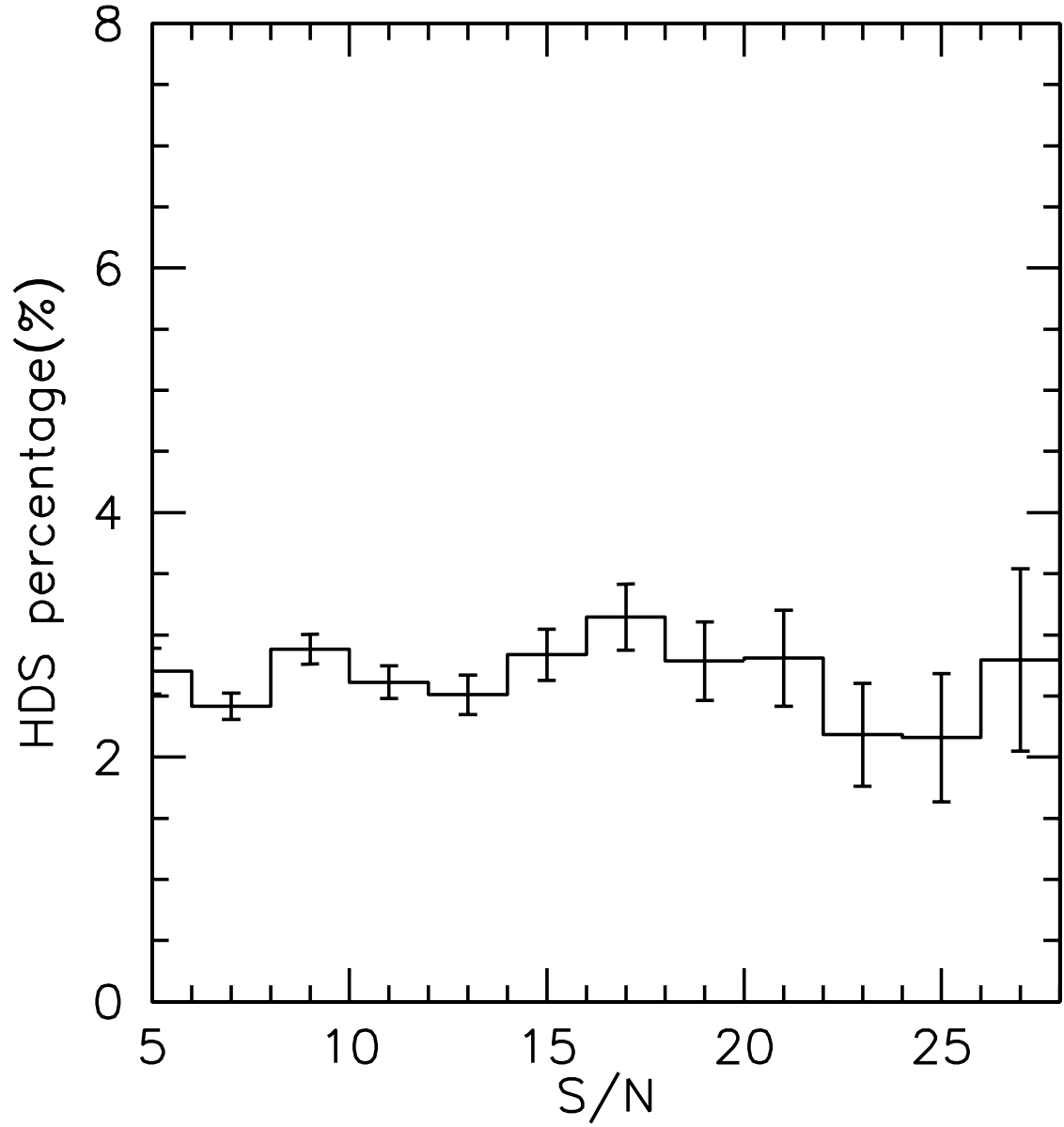}
\caption{
\label{fig:hds_sn}
The fraction of HDS galaxies as a function signal--to--noise in the $g$
band. The error bars are $\sqrt{N}$, where $N$ is the number of
galaxies in each bin.}
\end{figure}

In total, 3340 SDSS galaxies satisfy these criteria (Sample 1 plus Sample 2),
and we present these galaxies as our catalog of HDS galaxies. We note that
only 131 of these galaxies are securely identified as AGNs using the
prescription of Kewley et al. (2002) and Gomez et al. (2003). In Figure
\ref{fig:hds_sn}, we present the fraction of HDS galaxies selected as a
function of their signal--to--noise in SDSS $g$ band. It is re--assuring that
there is no observed correlation, which indicates that our selection technique
is not biased by the signal--to--noise of the original spectra.

For each galaxy in Samples 1 and 2, we present the unique SDSS Name (col.  1),
heliocentric redshift (col. 2), spectroscopic signal--to--noise in the SDSS
photometric $g$ band (col. 3), Right Ascension (J2000; col. 4) and Declination
(J2000; col. 5) in degrees, Right Ascension (J2000; col. 6) and Declination
(J2000; col. 7) in hours, minutes and seconds, the rest--frame EW(\hd) (\AA,
col. 8), the rest--frame $\Delta$EW(\hd) (\AA, col. 9), the rest--frame
EW(\oii) (\AA, col. 10), the rest--frame $\Delta$EW(\oii) (\AA, col. 11), the
rest--frame EW(\ha) (\AA, col. 12), the rest--frame $\Delta$EW(\ha) (\AA,
col. 13), the SDSS Petrosian $g$ band magnitude (col. 14), the SDSS Petrosian
$r$ band magnitude (col. 15), the SDSS Petrosian $i$ band magnitude (col. 16),
the SDSS Petrosian $z$ band magnitude (col. 17; all magnitudes are extinction
corrected), the k--corrected absolute magnitude in the SDSS $r$ band
(col. 18), SDSS measured seeing in $r$ band (col. 19), concentration index
(col. 20, see Shimasaku et al. 2001 and Strateva et al. 2001 for definition).
In Column 21, we present the AGN classification based on the line indices of
Kewley et al. (2002), and in Column 22, we present our E+A classification
flag, which is defined in Section \ref{previous}.  An electronic version of
our catalog can be obtained at {\tt
http://astrophysics.phys.cmu.edu/$\sim$tomo/ea}.$\footnote{Mirror sites
are available at http://kokki.phys.cmu.edu/$\sim$tomo/ea,\  http://sdss2.icrr.u-tokyo.ac.jp/$\sim$yohnis/ea}$

In addition to presenting Samples 1 and 2, we also present a volume--limited
sample selected from these two samples but within the redshift range of
$0.05<z<0.1$ and with M$(r^*)<-20.5$ (which corresponds to $r=17.7$ at
$z=0.1$, see Gomez et al. 2003).  We use Schlegel, Finkbeiner \& Davis (1998)
to correct for galactic extinction and Blanton et al. (2002b; v1\_11) to
calculate the k--corrections.  In Table \ref{tab:frequency}, we present the
percentage of HDS galaxies that satisfy our criteria. In this table, the
number of galaxies in the whole sample (shown in the denominator) changes
based on the number of galaxies that could have had their \hd, \oii\, and
\ha\, lines measured because of masked pixels in the spectra.

We note here that we have not corrected our sample for possible aperture
effects, except restrict the sample to $z\geq0.05$: A 3 arcsecond fiber
corresponds to $2.7h_{75}^{-1}$ kpc at this redshift, which is comparable to
the half--light radius of most our galaxies (see also Gomez et al. 2003).  We
see an increase of $0.33$\AA\, ($\simless10\%$) in the median observed \hd\ EW
for the whole HDS sample over the redshift range of our volume--limited sample
($0.05<z<0.1$). This is probably caused by more light from the disks of
galaxies going down the fiber at higher redshifts. We see no such trend for
the subsample of true E+A galaxies (see Section \ref{discussion}) in our HDS
sample.

\section{Discussion}
\label{discussion}

In this Section, we compare our sample of HDS galaxies against previous
samples of such galaxies in the literature. Further analysis of the global
properties (luminosity, environment, morphology, {\it etc.}) of these galaxies
will be presented in future papers.

\subsection{Comparison with Previous Work}
\label{previous}

In this Section, we present a preliminary comparison of our SDSS HDS galaxies
with other samples of post--starburst galaxies in the literature.  We have
attempted to replicate the selection criteria of these previous studies as
closely as possible, but in some cases, this is impossible. Furthermore, we
have not attempted to account for systematic differences in the distribution
of signal--to--noises, differences in the spectral resolutions, and
differences in the selection techniques used by different authors, {\it e.g.},
we are unable to fully replicate the criteria of Zabludoff et al. (1996) as we
do not yet possess accurate measurements of the H$\gamma$ and H$\beta$ lines.
Therefore, these are only crude comparisons and any small differences seen
between the samples should not be over--interpreted until a more detailed
analysis can be carried out. We summarize our comparison in Table
\ref{comparison} and discuss the details of these comparisons below: Table
\ref{comparison} does however, demonstrate once again the rarity of
\hd--strong galaxies, especially at low redshift, as well as illustrating the
sensitivity of their detection to the selection criteria used.

We first compare our sample to that of Zabludoff et al. (1996), which is the
most similar to our work, especially as the magnitude limits of the LCRS (used
by Zabludoff et al. 1996) and the SDSS are close, thus minimizing possible
bias.  Zabludoff et al. (1996) selected E+A galaxies from the LCRS using the
following criteria; a redshift range of $15,000 < cz < 40,000\, {\rm km\,
s^{-1}}$, a signal--to--noise of $>8$ per pixel, an EW of \oii\, of $>-2.5$\AA,
and an average EW for the three Balmer lines (\hd, H$\gamma$ and H$\beta$) of
$>5.5$\AA. Using these four criteria, Zabludoff et al. (1996) selected 21 LCRS
galaxies as E+A galaxies, which corresponds to 0.2\% of all LCRS that satisfy
the same signal--to--noise and redshift limits.  We find 80 SDSS galaxies
(from the whole SDSS dataset analyzed here) satisfy the same redshift range
and \oii\ EW detection limit as used by Zabludoff et al. (1996), as well as
having H$\delta$ EW of $>$5.5\AA, which should be close to the average of the
EW of the three Balmer lines (\hd, H$\gamma$ and H$\beta$) used by Zabludoff
et al. (1996). Of these 80 galaxies, 71 are in our HDS sample.  Given all
the caveats discussed above, it is re--assuring that we have found HDS
galaxies at a similar frequency (see Table \ref{comparison}) as Zabludoff et
al. (1996) and it suggests that our criteria are consistent.

Several other authors have used similar criteria to Zabludoff et al. (1996) to
search for post--starburst galaxies in higher redshift samples of galaxies.
For example, Fisher et al. (1998) used an average EW of the H$\delta$,
H$\gamma$ and H$\beta$ lines of $>4$\AA\ and an EW of \oii\, of $<5$\AA. With
these criteria, they found 4.7\% of their galaxies were E+A galaxies.
Similarly, Hammer et al. (1997) used an average of H$\delta$, H$\gamma$ and
H$\beta$ of $>5.5$\AA, an EW of \oii\, of $<$5--10 \AA\ and $M_B<$-20 to
select E+A galaxies from the CFRS field galaxy sample. They found that 5\% of
their galaxies satisfied these criteria.

We have also attempted to replicate the selection criteria of Dressler et
al. (1999) and Balogh et al. (1999) as closely as possible, using all SDSS
galaxies in the volume--limited sample with measured \hd, \oii\, and \ha, and
not simply the HDS subsample. For example, the MORPHS collaboration of
Dressler et al. (1999) and Poggianti et al. (1999) selected E+A galaxies using
a \hd\, EW of $>3$ \AA\ and an \oii\, EW of $<5$\AA\ . Using these criteria,
they found a significant excess of E+A galaxies in their 10 high redshift
clusters (21\%), compared with the field region (6\%). Alternatively, Balogh
et al. (1999) selected E+A galaxies using \hd\, EW of $>5$\AA\ and an \oii\,
EW of $<5$\AA\ . They found instead $1.5\pm0.8$\% of cluster galaxies were
classified as E+A galaxies compared to $1.2\pm0.8$\% for the field (brighter
than M$(r)=-18.8+5{\rm log}h$ after correcting for several systematic
effects).

In Table \ref{comparison}, we present a qualitative comparison of our HDS
sample with these higher redshift studies and, within the quoted errors, the
frequencies of HDS galaxies we observe are consistent with these
works. However, we caution the reader not to over--interpret these numbers for
several reasons. First, we are comparing a low redshift sample ($z<0.1$) of
HDS galaxies to high redshift studies ($z\simeq0.5$) of such galaxies, and we
have not accounted for possible evolutionary effects or observational
biases. In particular, we are comparing our sample against the corrected
numbers presented by Balogh et al. (1999), which attempt to account for
scatter in the tail of the \hd\, distribution due to the large intrinsic
errors on \hd\, measurements, while Dressler et al. (1999) do not make such a
correction.  Secondly, we are comparing a field sample of HDS galaxies to
predominantly cluster--selected samples of galaxies (were available, we only
quote in Table \ref{comparison} frequencies based on field samples). Finally,
the luminosity limit of our volume--limited sample is brighter than the high
redshift studies, which may account for some of the discrepancies.

\begin{figure}
\plotone{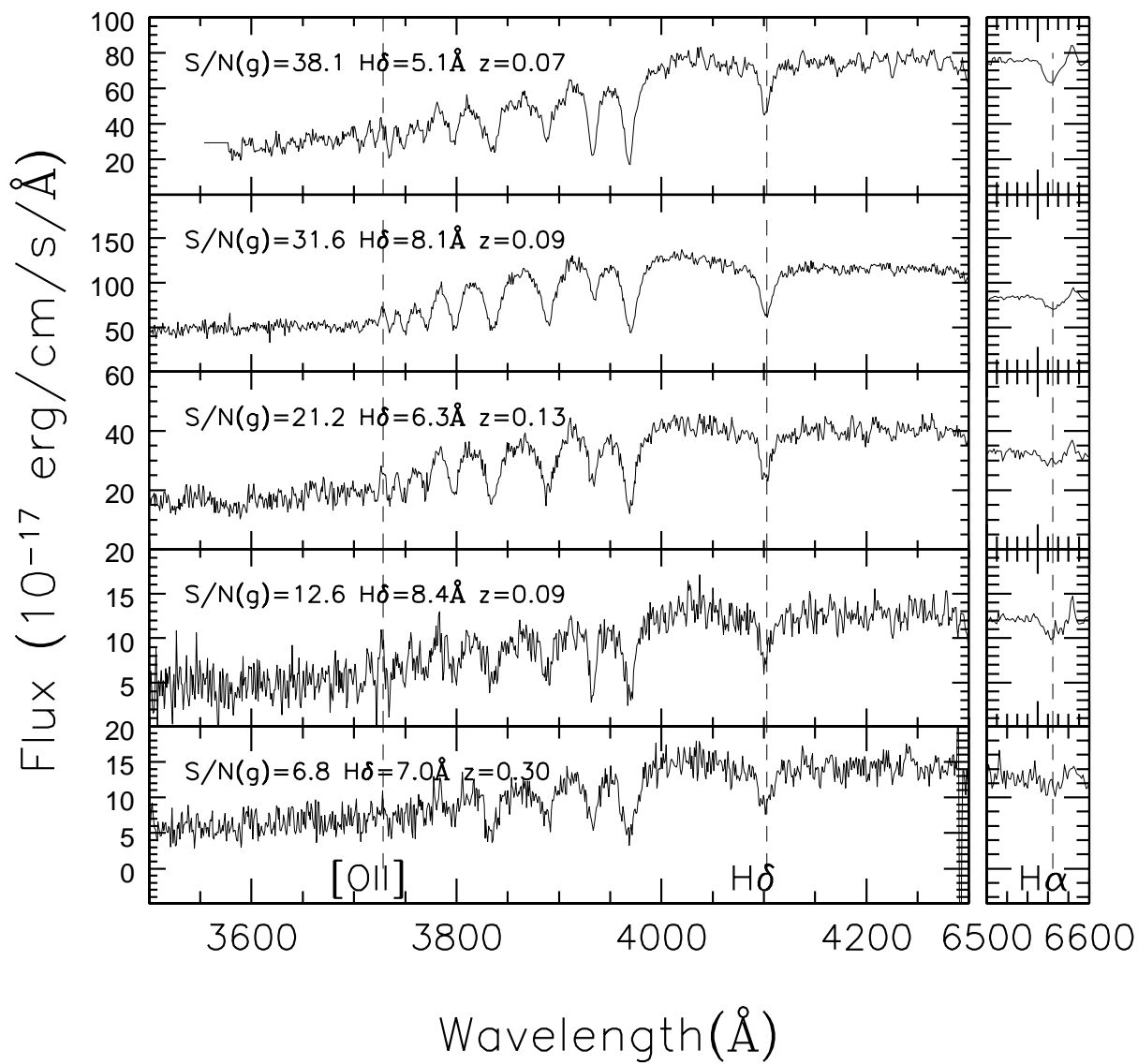}
\caption{
\label{fig:true_ea}
We show here five examples of spectra for the ``true E+A'' (k+a) subsample
of HDS galaxies discussed in Section \ref{discussion}.  These galaxies
possess strong Balmer absorption lines, but have no, or little,
detected \oii\, or \ha\, emission. We present a range of signal--to--noise
ratios, as well as provide the measured redshift and \hd\, EW for each galaxy.}
\end{figure}

Finally, we note that the original E+A phenomenon in galaxies, as discussed by
Dressler \& Gunn (1983, 1992), was defined to be a galaxy that possessed
strong Balmer absorption lines, but with no emission lines, {\it i.e.}, a
galaxy with the signature of recent star--formation activity (A stars), but no
indication of on--going star--formation ({\it e.g.}, nebular emission
lines). Given the quality of the SDSS spectra, we can re--visit this specific
definition and select galaxies from our sample that possess less than
$1\sigma$ detections of both the \ha\, and \oii\, emission lines, {\it i.e.},
${\rm EW([OII]) + \Delta EW([OII]) \geq0}$\AA\, and ${\rm EW(H\alpha) + \Delta
EW(H\alpha)\geq0}$\AA).  We find that only $3.5\pm0.7$\% (25/717) of galaxies
in the volume limited HDS sample satisfy such a strict criteria (see Table
\ref{tab:frequency}). We show example spectra of these galaxies in Figure
\ref{fig:true_ea} and highlight them in the catalog using the E+A
classification flag. This exercise demonstrates that true E+A galaxies -- with
no, or little, evidence for on--going star--formation -- are extremely rare at
low redshift in the field, {\it i.e.}, $0.09\pm0.02$\% of all SDSS galaxies in our volume--limited sample.

\subsection{HDS Galaxies with Emission--lines}

\begin{figure}
\plotone{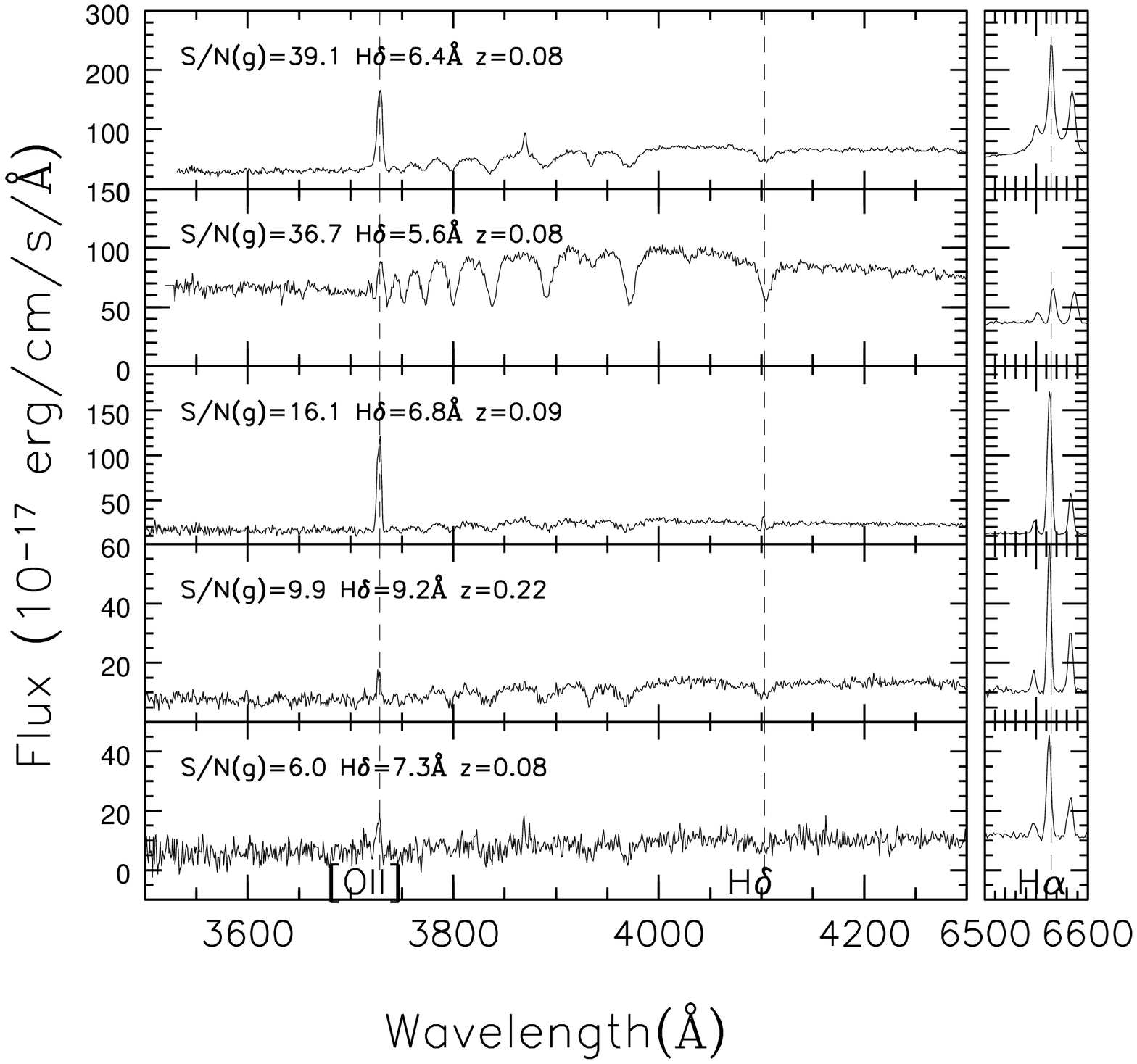}
\caption{
\label{fig:ea_em}
We present here five example spectra of our HDS galaxies that
possess detected \oii\, and \ha\, emission lines. 
We present a range of signal--to--noise
ratios, as well as provide the measured redshift and \hd\, EW for each galaxy.}
\end{figure}

In this Section, we examine the frequency of nebular emission lines (\oii,
\ha) in the spectra of our HDS galaxies. This is possible because of the large
spectral coverage of the SDSS spectrographs which allow us to study both the
\oii\, and \ha\, emission lines for all galaxies out to a redshift of
$\simeq0.35$.  We begin by looking at HDS galaxies that possess both the \ha\,
and \oii\, emission lines.  Using the criteria that both \oii\, and \ha\,
lines must be detected at $>1\sigma$ significance ({\it i.e.}, ${\rm EW([OII])
+ \Delta EW([OII]) <0}$\AA\, and ${\rm EW(H\alpha) + \Delta EW(H\alpha)<0}$\AA),
we find that $89\pm5$\% (643/717) of HDS galaxies in our volume--limited
sample are selected. Of these, 131 HDS galaxies possess a robust detection of
an AGN, based on the line indices of Kewley et al. (2001), similar to the AGN
plus post--starburst galaxy found recently in the 2dFGRS (see Sadler, Jackson
\& Cannon 2002). Therefore, a majority of these emission line \hd\, galaxies
may have on--going star formation and are similar to the e(a) and A+em
subsample of galaxies discussed by Poggianti et al. (1999) and Balogh et
al. (1999). We show in Figure \ref{fig:ea_em} examples of these HDS galaxies
that possess both the \oii\, and \ha\, emission lines.  The median SFR of
these galaxies (calculated from \ha\, flux, see Kennicut 1998) is
$\simeq0.5{\rm M_{\odot}/yr}$, with a maximum observed SFR of $50{\rm
M_{\odot}/yr}$. We note that these SFRs have not been corrected for dust
extinction or aperture effects, which could result in them being a factor of
$5$ to $10$ lower than the true SFR for the whole galaxy (see Hopkins et
al. in prep).

\begin{figure}
\plotone{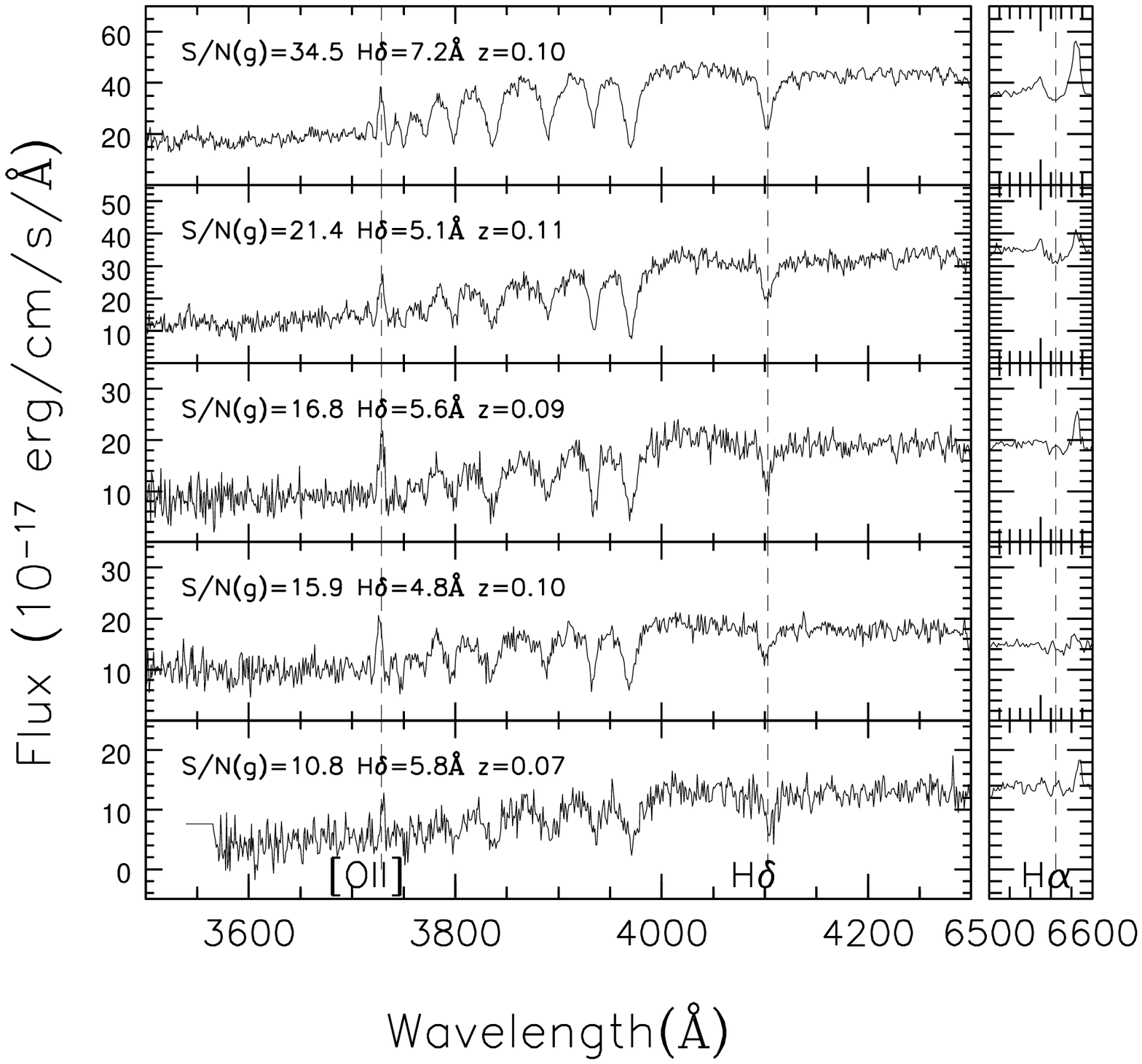}
\caption{
\label{fig:oii}
We present here five example spectra of our HDS galaxies that
possess detected \oii\, emission lines, but no detected \ha. We
present a range of signal--to--noise ratios, as well as provide the
measured redshift, \hd\, EW and name for each galaxy.}
\end{figure}

We next examine the frequency of HDS galaxies with \oii\, emission, but no
detectable \ha\, emission. Using the criteria of EW(\ha) $<$1 $\sigma$
detection and EW(\oii) $>$ 1$\sigma$ detection ({\it i.e.}, ${\rm
EW([OII]) + \Delta EW([OII]) <0}$\AA\, and ${\rm EW(H\alpha) + \Delta EW(H\alpha)
\geq0}$\AA), we find $2.9\pm0.65$\% (21/717) of our HDS galaxies in the
volume--limited sample satisfy these cuts. The presence of \oii\, demonstrates
that the galaxy may possess on--going star formation activity, yet the lack of
the \ha\, emission is curious. Possible explanations for this phenomena are
strong self-absorption of \ha\, by the many A stars in the galaxy and/or
metallicity effect which could increase the \oii\, emission relative to \ha\,
emission. We show several examples of these galaxies in Figure \ref{fig:oii},
and the lack of \ha\, emission is clearly visible. The fact that many of these
galaxies possess strong \nii\, lines (flanking the \ha\, line) indicates
strong self--absorption is a likely explanation for the missing \ha\, emission
line. Median [OII] EW of these galaxies is 1.3 \AA. Compared with 11.5
\AA of HDS galaxies with both [OII] and H$\alpha$ emission, these
galaxies have much lower amount of [OII] in emission.

Finally, we find that $3.8\pm 0.7$\% (27/717) of our HDS galaxies in our
volume--limited sample satisfy the following criteria; ${\rm EW([OII]) +
\Delta EW([OII]) \geq0}$\AA\, and ${\rm EW(H\alpha) + \Delta EW(H\alpha)
<0}$\AA. In comparison, only 52 of our HDS galaxies, the volume limited
sample, have just no \oii\ emission detected (only EW([OII]) + $\Delta$
EW([OII]) $\geq$0\AA). Therefore, 52$\pm$12\% (27/52) of the galaxies with no
detected \oii, have detected \ha\ emission.  The existence of such galaxies
has ramifications on high redshift studies of post--starburst galaxies, as
such studies use the \oii\, line to constrain the amount of on--going
star--formation within the galaxies {\it e.g.}, Balogh et al. (1999),
Poggianti et al. (1999).  Therefore, if the \ha\, emission comes from
star--formation activity, then these previous high redshift studies of post
starburst galaxies may be contaminated by such galaxies.  A possible
explanation for the lack of \oii\, emission is dust
extinction {\it e.g.}, Miller \& Owen (2002) recently
found dusty star--forming galaxies which do not possess \oii\, in emission, but
have radio fluxes consistent with on--going star formation activity. This
explanation would also be consistent with the low signal--to--noise we observe
in the blue--end of the SDSS spectra of these galaxies, relative to the
signal--to--noise seen in the red--end of their spectra. Median H$\alpha
$ EW of these galaxies are 1.5\AA, whereas that of HDS galaxies with
both [OII] and H$\alpha$ emission is 25.9\AA.

\section{Conclusions}
\label{conclusion}

We present in this paper the largest, most homogeneous, search yet for
\hd--strong galaxies ({\it i.e.}, post--starburst galaxies, E+A's, k+a's,
a+k's {\it etc.}) in the local universe. We provide the astronomical community
with a new catalog of such galaxies selected from the Sloan Digital Sky Survey
(SDSS) based solely on the observed strength of the \hd\, Hydrogen Balmer
absorption line within the spectrum of the galaxy. We have carefully studied
different methodologies of measuring this weak absorption line and conclude
that a non--parametric flux--summing technique is most suited for an automated
application to large datasets like the SDSS, and more robust for the observed
signal--to--noise ratios available in these SDSS spectra.  We have studied the effects of
dust extinction, emission--filling and stellar absorption upon the measurements
of our \hd\, lines and have determined the external error on our measurements,
as a function of signal--to--noise, using duplicate observations of 11358
galaxies in the SDSS. In total, our catalog of \hd--selected (HDS) galaxies
contains 3340 of the 95479 galaxies in the Sloan Digital Sky Survey (at the
time of writing) and we present these HDS galaxies to the community to help in
the understanding of such systems and to aid in the comparison with higher
redshift studies of post--starburst galaxies.

The measured abundance of our \hd--selected (HDS) galaxies is $2.6\pm$0.1\% of
all galaxies within a volume--limited sample of $0.05<z<0.1$ and
M($r^*$)$<-20.5$, which is consistent with previous studies of post--starburst
galaxies in the literature. We find that only 25 galaxies ($3.5\pm0.7$\%) of
HDS galaxies in this volume limited sample show no, or little, evidence for
\oii\, and \ha\, emission lines, thus indicating that true E+A galaxies (as
originally defined by Dressler \& Gunn) are extremely rare objects, {\it
i.e.}, only $0.09\pm0.02$\% of all galaxies in our volume--limited sample. In
contrast, $89\pm5$\% of our HDS galaxies have significant detections of the
\oii\, and \ha\, emission lines. Of these, 131 galaxies are robustly
classified as Active Galactic Nuclei (AGNs) and therefore, a majority of these
emission line HDS galaxies are star--forming galaxies, similar to the e(a) and
A+em galaxies discussed by Poggianti et al. (1999) and Balogh et al. (1999).
We will study the global properties of our HDS galaxies in further detail in
future papers.

\acknowledgments

We are indebted to Michael Balogh and Ann Zabludoff who were instrumental in
the design and construction of this sample and have helped us understand the
previous literature on this subject. We also thank Ian Smail and Bianca
Poggianti for their enthusiasm for this work and their many discussions about
post--starburst galaxies. We thank Percy Gomez and Michael Strauss for their
help throughout this project.  RCN \& TG thanks the Physics Department of
Durham University for their hospitality during the Summer of 2002, when part
of this work was carried out. T. G. acknowledges financial support from the
Japan Society for the Promotion of Science (JSPS) through JSPS Research
Fellowships for Young Scientists.  Funding for the creation and distribution
of the SDSS Archive has been provided by the Alfred P. Sloan Foundation, the
Participating Institutions, the National Aeronautics and Space Administration,
the National Science Foundation, the U.S. Department of Energy, the Japanese
Monbukagakusho, and the Max Planck Society. The SDSS Web site is
http://www.sdss.org/.  The SDSS is managed by the Astrophysical Research
Consortium (ARC) for the Participating Institutions. The Participating
Institutions are The University of Chicago, Fermilab, the Institute for
Advanced Study, the Japan Participation Group, The Johns Hopkins University,
Los Alamos National Laboratory, the Max-Planck-Institute for Astronomy (MPIA),
the Max-Planck-Institute for Astrophysics (MPA), New Mexico State University,
University of Pittsburgh, Princeton University, the United States Naval
Observatory, and the University of Washington.

\clearpage

\begin{table}[h]
\begin{center}
\caption{
\label{tab:wavelength}
The wavelength ranges used to measure our \hd, \oii\, and \ha\, EWs.
}
\begin{tabular}{llll}
\hline
  & Blue continuum  & Line & Red continuum \\
\hline
\hline
H$\delta$ (narrow) &  4030-4082\AA &    4088-4016\AA & 4122-4170\AA  \\
H$\delta$ (wide)   &  4030-4082\AA &    4082-4022\AA & 4122-4170\AA  \\ 
 \oii            &  3653-3713\AA &    3713-3741\AA & 3741-3801\AA  \\ 
H$\alpha$          &  6490-6537\AA &    6555-6575\AA & 6594-6640\AA  \\ 
\hline
\end{tabular}
\end{center}
\end{table}

\begin{table}[h]
\begin{center}
\caption{
\label{equation} Coefficients of third order polynomial fits to the error distributions shown in Figures \ref{fig:err_from_double_obs_hd},
\ref{fig:err_from_double_obs_oii} and \ref{fig:err_from_double_obs_ha}.
}
\begin{tabular}{lllll}
\hline Line & a$_0$ & a$_1$ & a$_2$ & a$_3$  \\ \hline \hline
 H$\delta$ &  2.98 & -0.28 & 0.012 &   -0.00018\\
 \oii &    4.96 &  -0.39 &     0.014 & -0.00016\\
 H$\alpha$ &  3.74 & -0.36 &  0.014 &  -0.00017\\
\hline \hline
\end{tabular}
\end{center}
\end{table}

\begin{table}[h]
\begin{center}
\caption{
\label{tab:frequency}
The frequency of finding HDS galaxies.}
\begin{tabular}{llll}
\hline
 Category & \% (All galaxies) & \% (Volume Limited) \\
\hline
\hline
Whole HDS sample  & 3340/95479 (3.50$\pm$0.06\%)  & 717/27014 (2.6$\pm$0.1\%) \\
True ``E+A''      & 140/94770 (0.15$\pm$0.01\%)   & 25/26863 (0.09$\pm$0.02\%) \\ 
\hline
\end{tabular}
\end{center}
\end{table}

\begin{table}[h]
\begin{center}
\caption{
\label{comparison}
A comparison of our HDS sample of galaxies to previous work in the literature.
 }
\begin{tabular}{lllll}
\hline Author & Balmer lines & Emission & Their \% (field) & Our \% \\ \hline \hline
Zabludoff et al. & H$\delta>$5.5\AA & [OII]$>$-2.5\AA
& 0.19$\pm$0.04\% & 0.16$\pm$0.02\% (80/49994)  \\ 
Poggianti et al. & H$\delta>$3 \AA & [OII]$>$-5 \AA & 6$\pm$3\% & 5.79$\pm$0.15\% (1565/27014) \\ 
Balogh et al. & H$\delta>$5 \AA & [OII]$>$-5 \AA & $1.2\pm0.8$\%  & 0.74$\pm$0.05\% (200/27014) \\
\hline \hline
\end{tabular}
\end{center}
\end{table}

\end{document}